\documentclass{vgtc}                          

\ifpdf
  \pdfoutput=1\relax                   
  \usepackage{graphicx}                
  \DeclareGraphicsExtensions{.pdf,.png,.jpg,.jpeg} 
\else
  \usepackage{graphicx}                
  \DeclareGraphicsExtensions{.eps}     
\fi%

\usepackage{microtype}                 
\PassOptionsToPackage{warn}{textcomp}  
\usepackage{textcomp}                  
\usepackage{mathptmx}                  
\usepackage{times}                     
\usepackage{cite}                      
\usepackage{tabu}                      
\usepackage{booktabs}                  

\usepackage{float}
\usepackage[labelfont=it,textfont={bf,it}]{caption}
\usepackage{caption}
\usepackage{subcaption}
\usepackage{url}
\usepackage{paralist}
\usepackage{todonotes}
\usepackage[pdfpagelabels,colorlinks,allcolors=blue]{hyperref}
\usepackage{xspace}

\usepackage{amsmath}

\graphicspath{{Figures/}}

\onlineid{8231}

\vgtccategory{Techniques}

\preprinttext{To appear in IEEE PacificVis.}

\title{Browser-based Hyperbolic Visualization of Graphs}

\author{Jacob Miller\thanks{e-mail: jacobmiller1@email.arizona.edu}\\ %
        \scriptsize University of Arizona %
\and Stephen Kobourov\thanks{e-mail: kobourov@cs.arizona.edu}\\ %
     \scriptsize University of Arizona %
\and Vahan Huroyan\thanks{e-mail: vahanhuroyan@math.arizona.edu}\\    \scriptsize University of Arizona
 }

\teaser{
  \centering
  \includegraphics[width=0.32\textwidth]{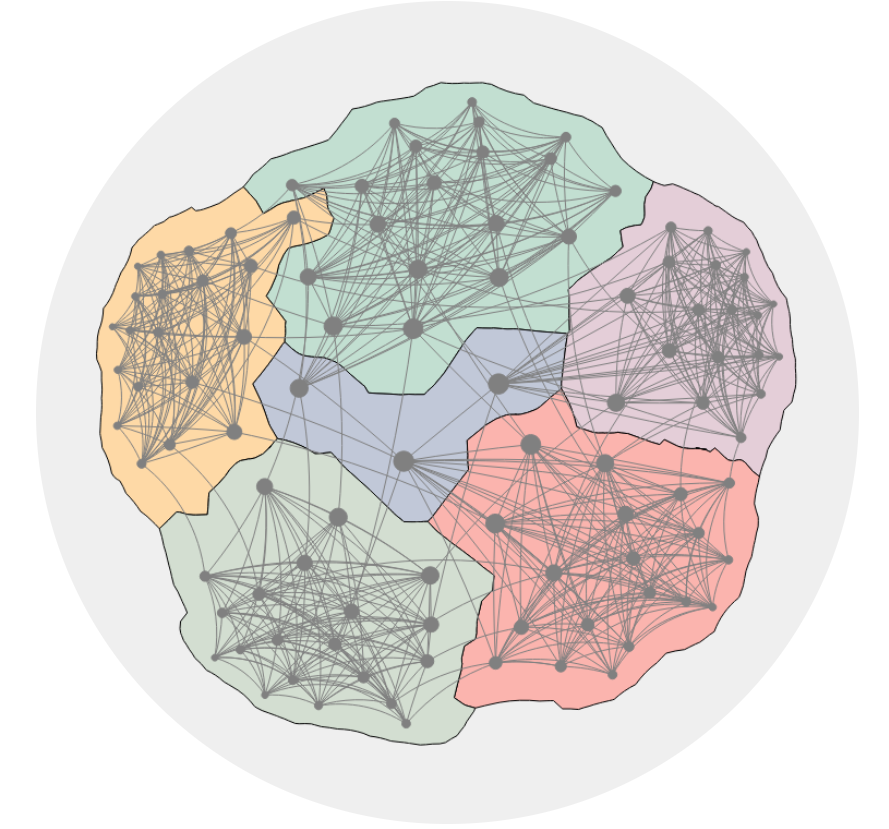}
  \includegraphics[width=0.32\textwidth]{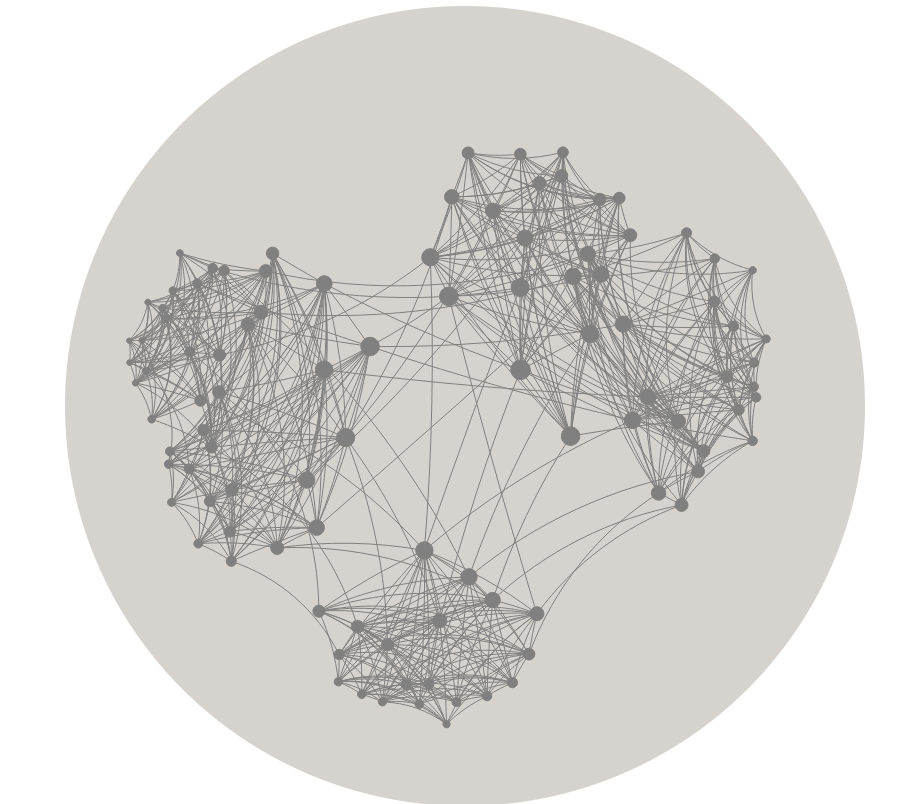}
  \includegraphics[width=0.32\textwidth]{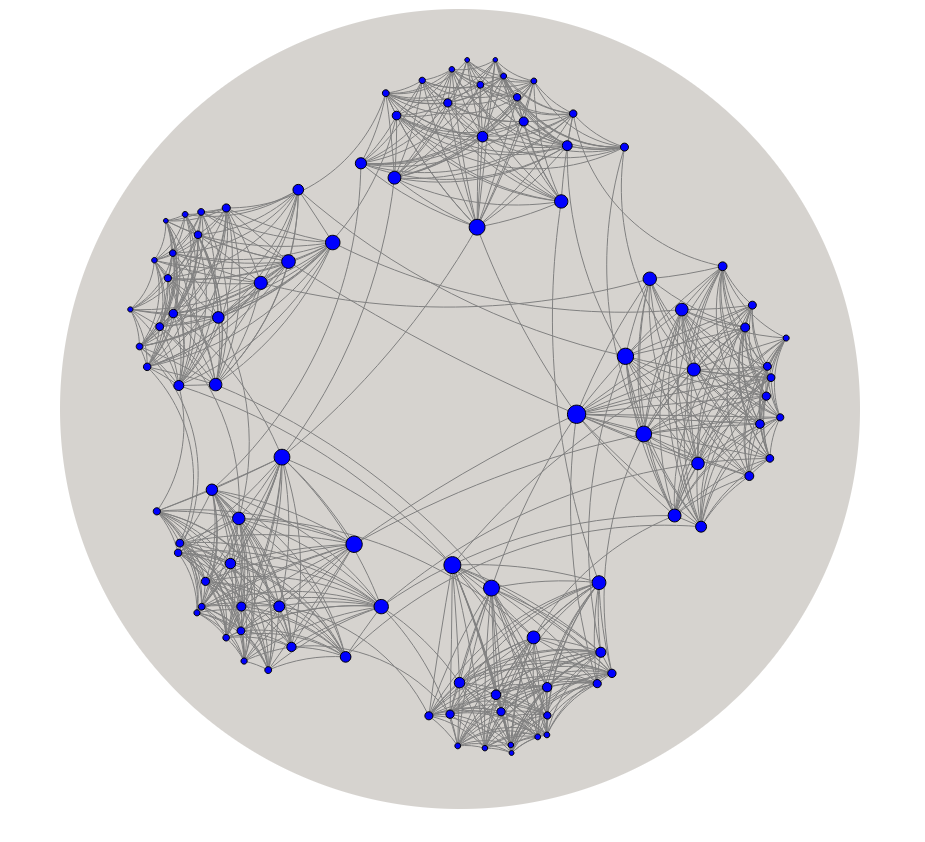}
  \caption{Example layouts of the same graph, generated by the three hyperbolic graph embedding algorithms discussed in this paper: inverse projection (left), force-directed (center) and hyperbolic-MDS (right).} 
  \label{fig:teaser}
}

\abstract{  

Hyperbolic geometry offers a natural `focus+context' for data visualization and has been shown to underlie real-world complex networks. However, current hyperbolic network visualization approaches are limited to special types of networks and do not scale to large datasets.
With this in mind, we designed, implemented, and analyzed three methods for hyperbolic visualization of networks in the browser based on inverse projections, generalized force-directed algorithms, and hyperbolic multi-dimensional scaling (H-MDS). 
A comparison with Euclidean MDS shows that H-MDS produces embeddings with lower distortion for several types of networks.
All three methods can handle node-link 
representations and are available in fully functional web-based systems.

} 

\CCScatlist{
  \CCScatTwelve{Graph drawing}{}{}{};
  \CCScatTwelve{Hyperbolic geometry}{}{}{};
  \CCScatTwelve{Non-Euclidean embedding}{}{}{};
  \CCScatTwelve{Stochastic gradient descent}{}{}{}
}

\begin{document}


\firstsection{Introduction}

\maketitle

Node-link representations of graphs in the 2-dimensional Euclidean plane are the most typically used graph visualizations.
The structure of many graphs, notably planar graphs, can be realized well in the plane, but others are better represented in non-Euclidean geometries. For example, 3-dimensional polytopes are well represented in spherical space, while large hierarchies such as trees can be cleanly embedded in hyperbolic space. Standard hyperbolic projections into Euclidean space also provide
a natural `focus+context' view of the graph, with parts of the graph  near the center of the view shown  large and those far from the center progressively smaller, with the entire graph being in the view. 

A recent work~\cite{krioukov2010hyperbolic} suggests that hyperbolic geometry underlies complex networks, in a similar way as spherical geometry underlies geographic data.


Though there has been some work
on visualizing hierarchies
using hyperbolic space in the browser~\cite{Michael2018survey}, there are no tools that support browser-based hyperbolic visualization of
general graphs.

We describe three methods for laying out graphs in the 2-dimensional hyperbolic space, $H^2$. The first method relies on taking a
pre-computed Euclidean layout of a graph and projecting it into hyperbolic space, providing standard map interactions,
such as pan, zoom, re-center, click and drag. We implement this method in a web based system that provides several layout algorithms for node-link and map-based visualization. This allows us to view and interact with GMaps, MapSets, BubbleSets, and LineSets in hyperbolic space.
The second method makes use of a generalization of force-directed algorithms to Riemannian geometries~\cite{kobourov2005non}.
We exploit the locally Euclidean properties of hyperbolic space so that  with the help of M\"obius transformations we can accurately model the forces.
In particular, this approach allows us to compute layouts where distances between nodes in hyperbolic space correspond to the underlying graph-theoretic distances between them. 
%
The third method attempts to directly realize graph distances in $H^2$ through a hyperbolic generalization of multidimensional scaling (MDS)~\cite{walter2002interactive,pd-mds}. For this method we adapt stochastic gradient descent (SGD) to hyperbolic space, as SGD has been shown to be efficient and produce high-quality layouts in Euclidean space~\cite{DBLP:journals/tvcg/ZhengPG19}.  To the best of our knowledge, there are no prior methods to adapt Euclidean layouts  to hyperbolic space, nor any hyperbolic SGD approaches. All three methods are available online. The projection method is available through GMap at {\small \url{http://gmap.cs.arizona.edu}}. The other two methods are available as a webapp on GitHub at {\small\url{https://github.com/Mickey253/hyperbolic-space-graphs}}.


\section{Related Work}

The graph layout problem typically involves placing nodes and routing edges in 2-dimensional
Euclidean space. Force-directed algorithms model the
system as a set of springs and attempt to balance the forces on nodes. Both their conceptual simplicity
and their generally aesthetically pleasing results have made this class of algorithms particularly useful for computing
graph layouts~\cite{kobourov2012spring}. Force-directed algorithms have been generalized to Riemannian geometries, (e.g., spherical and hyperbolic) by computing tangent planes at each node~\cite{kobourov2005non}. 

To the best of our knowledge, there are no browser-based tools for visualizing general graphs in hyperbolic space. Table~\ref{tab:overview} gives an overview of previous work in hyperbolic network visualization.
One of the
earliest approaches by Lamping  {\it et al.}~\cite{lamping1995} embeds hierarchies into the hyperbolic plane
by recursively placing each node's children evenly spaced around the arc of a circle.
This is possible thanks to the exponential expansion intrinsic to the geometry. 
They make use of the Poincar\'e projection to display the graph on
the computer monitor, which also provides the now well known `focus+context' effect. Navigating the hierarchy is done by re-centering the projection at a new
node in the hyperbolic plane. The embedding can be computed in linear time and arbitrary graphs can also be visualized using this approach by utilizing a spanning tree of the graph and `filling in' the rest of the edges later.
\begin{table}[h]
 \caption{Hyperbolic browsing systems}\vspace{1ex} 
 \label{tab:overview}
 \scriptsize
 \centering 
   \begin{tabular}{|r|r|r|r}
   \hline
     System & Date  & Description\\
     \hline
     \small{H2 Tree Browser} & 1995  & 2 dimensions, hierarchy viewer\\
     HVS & 2007 & Hiearchy visualization application\\
     Js InfoVis Toolkit & 2013 & Web-based data vis suite\\
     Treebolic & 2014 & Java Hyperbolic Poincar\'e visualization\\
     d3-hypertree & 2018 & Hyperbolic tree visualization library\\
     \hline
     H3 & 2000  & 3 dimensions, hierarchy viewer\\
     walrus & 2000  & Re-implementation of H3 in Java\\
     h3py & 2015  & Re-implementation of H3 in Python\\
     \hline
   \end{tabular}
\end{table}


A bioinformatics-motivated java application by Bingham and Sudarsanam~\cite{bingham2000visualizing} uses a similar approach to visualize  phylogenetic trees.
Andrews {\it et al.}~\cite{HVS} also rely on Lamping {\it et al.}'s work
in their Hierarchy Visualization System as do Baumgartner and Waugh~\cite{baumgartner2002roget2000} who visualize Roget's thesaurus. The Java InfoVis Toolkit also implements a hyperbolic hierarchy browser~\cite{jit} and TreeBolic implements the hyperbolic  tree layout~\cite{treebolic}.
More recently, Glatzhofer developed a hyperbolic hierarchy browser utilizing d3.js, a javascript graphics library which works in the browser, and can display large hierarchies smoothly with different layout algorithms~\cite{hyperbolicTOL,Michael2018survey,d3-hypertree}.

While most prior work considers the 2-dimensional hyperbolic plane, Munzner has also used 3D hyperbolic space to visualize hierarchies with the help of the Beltrami-Klein projection~\cite{munzner1997h3,munzner1998exploring,munzner2000interactive,munzner1995visualizing}.
Here geodesics are mapped to straight lines rather than the circular arcs of the Poincar\'e projection. 
Munzner's work has been re-implemented in two subsequent systems: Walrus~\cite{walrus2000} and h3py~\cite{h3py}.

Hyperbolic space has been explored in the context of non-linear dimensionality reduction, specifically multi-dimensional
scaling (MDS). The idea is to match pairwise similarities with distances in an embedding:
the more similar two elements are, the closer they are in the embedding. The Euclidean distance is traditionally
used as a closeness metric. Computing a graph layout can be interpreted as an MDS problem by treating the graph theoretic distance between pairs of nodes as their pairwise distance metric. 
%

There are three different types of MDS: classical, metric, and non-metric (although these labels are not used consistently in different fields)
Here, we refer to Torgerson's MDS as classical MDS, where the input distances are converted to similarities, and principal component analysis (PCA) is used to obtain the embedding~\cite{torgerson1952multidimensional}. 
Metric MDS  minimizes a loss function, commonly known as \textit{stress}~\cite{shepard1962analysis}. 
Finally, in non-metric MDS the input distances are not necessarily distances, but can be ranks~\cite{kruskal1964multidimensional}.

Classical MDS has been explored in hyperbolic space, by replacing the conversion to similarities with an appropriate hyperbolic scaling function~\cite{Clough2018,sala2018representation}. Using a similar idea, metric and non-metric MDS have been generalized to hyperbolic space by incorporating hyperbolic  geodesic distance into the cost function~\cite{walter2003interactive,walter2004hmds,walter2002interactive,zhou2021hyperbolic}.

It has been shown that some graphs can 
be embedded with lower error in hyperbolic space than in  Euclidean space~\cite{blasius2018efficient}. Zhou and Sharpee~\cite{zhou2021hyperbolic} show that hyperbolic MDS (H-MDS) can be used to detect the underlying geometry of a dataset, when comparing its embedding error to Euclidean non-metric MDS. They go further to show that the underlying space of genomes is hyperbolic. Krioukov {\it et al.}'s~\cite{krioukov2010hyperbolic} work indicates that hyperbolic geometry may underlie complex networks and hierarchical networks, such as phylogenetic trees and the internet. 

Greedy embeddings also appear to have a close relationship with hyperbolic geometry. Indeed, any connected, finite graph admits a greedy embedding in hyperbolic space, which is not generally true in Euclidean geometry~\cite{Kleinberg07}. Greedy embeddings of graphs allow for greedy routing, which is particularly useful when a node may not know the global topology, but only its own position and that of its neighbors such as in social networks and the internet~\cite{DBLP:journals/tc/EppsteinG11}. 

An open-source hyperbolic visualization tool, RogueViz~\cite{celinska2017programming}, includes different projections and educational tools, although its restriction to tessellations of the hyperbolic plane makes it less than ideal for general graphs. 
Self-organizing maps have been generalized to hyperbolic space, but are restricted to lattices~\cite{ontrup2001hyperbolic}.

Stress-based  approaches have been explored in other Riemannian spaces such as the sphere~\cite{perry2020drawing,DBLP:conf/chi/DuCLXT17} and the torus~\cite{DBLP:conf/chi/ChenDMB20,DBLP:conf/chi/ChenDBM21}, as different spaces  provide different visualization advantages. Unlike in the plane, on the sphere one can avoid issues such as central/peripheral placement, 
and on the torus larger classes of graphs can be drawn without crossings. Human subject studies show that these spaces are no worse than Euclidean space for common navigation tasks~\cite{DBLP:conf/chi/DuCLXT17,DBLP:conf/chi/ChenDMB20}.

One can achieve a similar focus+context effect by using lens effects~\cite{tominski2017interactive,10.2312:eurovisstar.20141172}. In particular, at first glance the 
Poincar\'e disk appears to resemble a fisheye lens. However, a lens effect generally applies only to a subset of the visible data, scaling or warping it to bring it into focus. 
The focus+context view is applied across all of the data in the Poincar\'e disk, with an exponential decrease in data size away from the center, but the entirety of the object remaining in view. 

It is worth mentioning that there are also theoretical limits on the effectiveness of hyperbolic embeddings for general graphs. Some graphs can be embedded trivially with a low, constant embedding error (e.g., as cycles and square lattices) but have non-trivial embedding error in the hyperbolic plane~\cite{eppstein2021,verbeek2016}. 
However, other graphs such as trees and hyperbolic tilings can be embedded better in hyperbolic space than in Euclidean space. For example, while Euclidean geometry only admits 3 regular tessellations (triangles, squares, hexagons), the hyperbolic plane admits infinitely many.

Zheng {\it et al.} show that stochastic gradient descent (SGD) can be used  effectively to solve MDS for graph layout in Euclidean space~\cite{DBLP:journals/tvcg/ZhengPG19}. In this paper we show  that SGD can also be deployed in hyperbolic space and produces good embeddings.

\begin{figure}
\centering
  \includegraphics[width=0.46\linewidth]{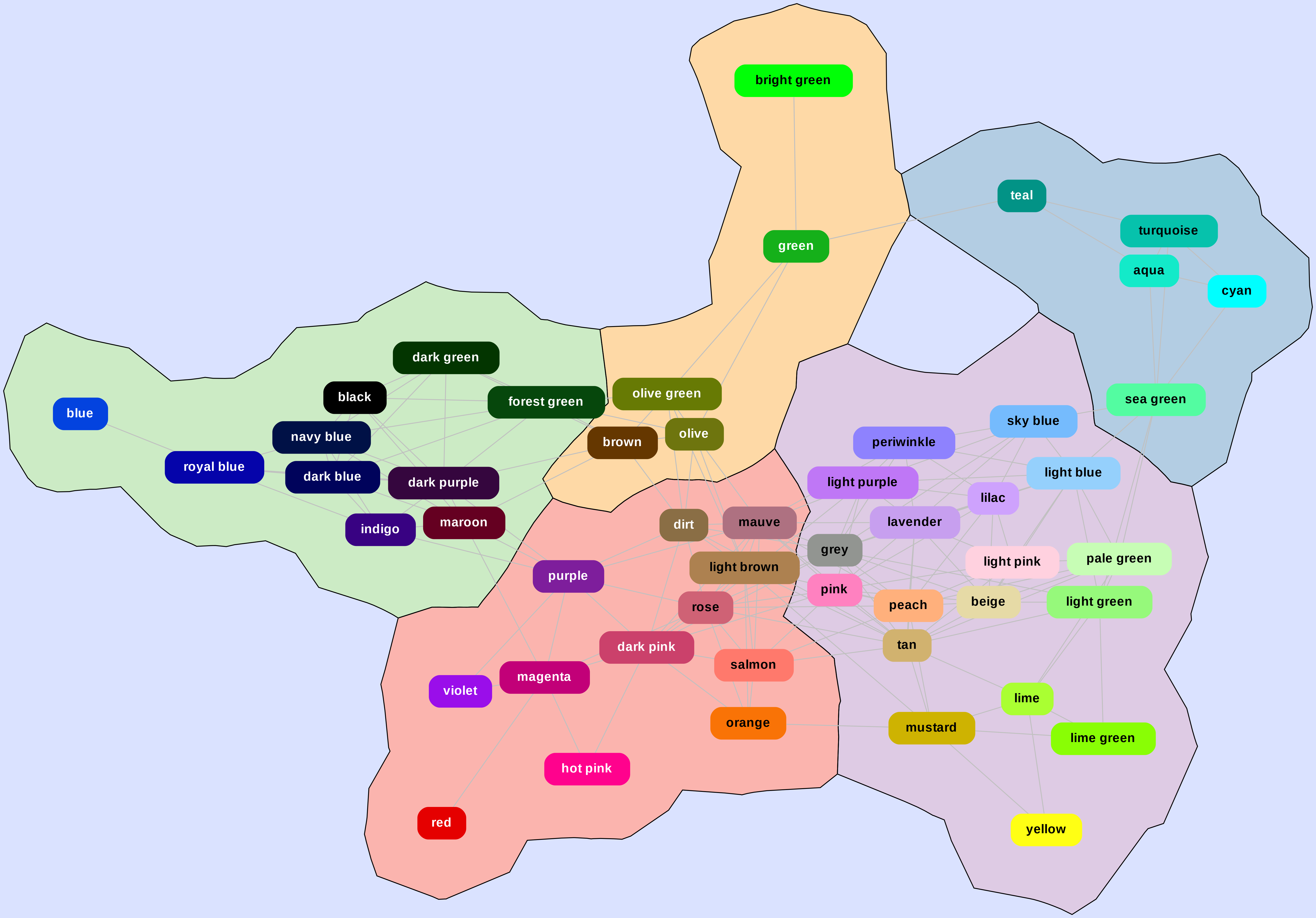}
  \includegraphics[width=0.50\linewidth]{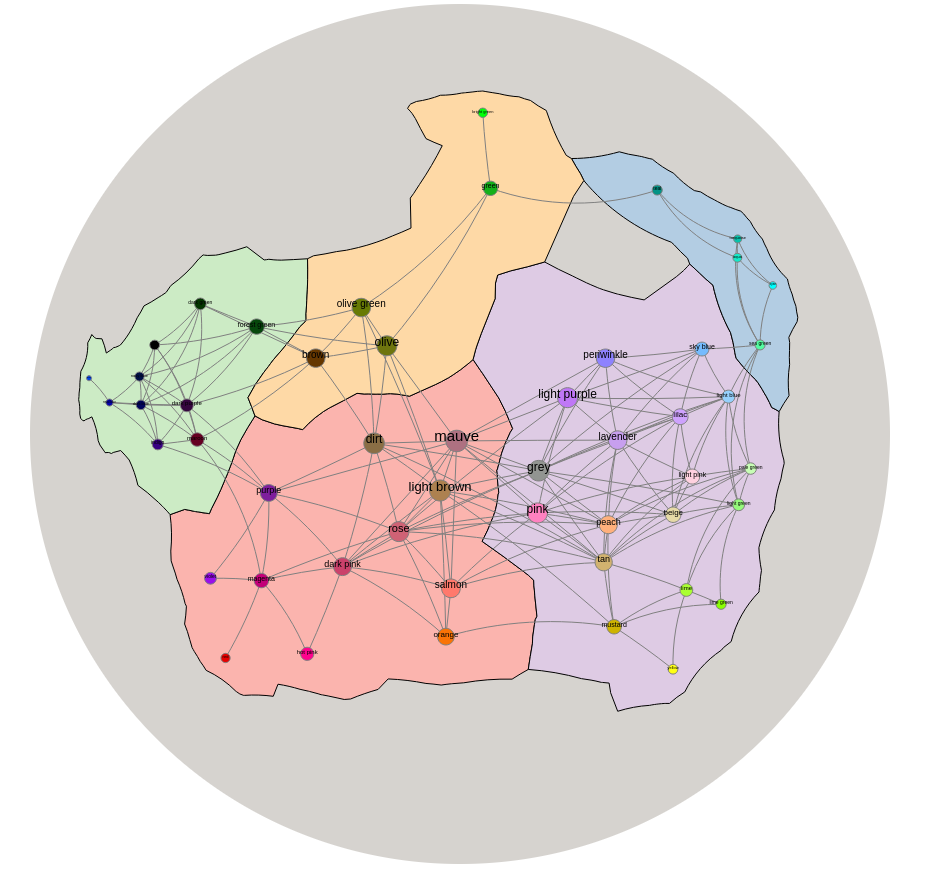}
\caption{An example of a GMap Euclidean layout (left) and its hyperbolic realization (right) via inverse projection.}
\label{fig:hyper_realized}
\end{figure}

\section{Projection-based Method}
The first method we present is based on the idea of starting with a precomputed layout and projecting it to the hyperbolic plane.
The implementation is available on the web, 
in a browser based graph visualization system. 
The system offers several layout algorithms, clustering algorithms and visualization styles, with a focus on map-like representations such as GMaps~\cite{gansner2010gmap}, MapSets~\cite{efrat2014mapsets}, BubbleSets~\cite{collins2009bubble}, and LineSets~\cite{alper2011design}.
  Several human-subject studies suggest that such map-like visualizations are at least as good as traditional node-link diagrams when it comes to task performance, memorization, and recall of the data~\cite{saket2015map,saket2014node}.

The Euclidean layouts are computed then saved in the graphviz DOT file format which includes graph-wide attributes, a node list, and an adjacency list~\cite{DBLP:conf/gd/EllsonGKNW00}. 
GMap computes the layout and stores node positions
as Cartesian coordinates. This is sufficient to draw node-link diagrams. Polygons given as a set of vertices
are stored as a graph-wide attribute along with their colors for the other map-like
layouts: GMaps, MapSets, BubbleSets and LineSets. Parsing the polygons is done as in~\cite{perry2020drawing}.

\begin{figure}
    \centering
        \includegraphics[width=.48\linewidth]{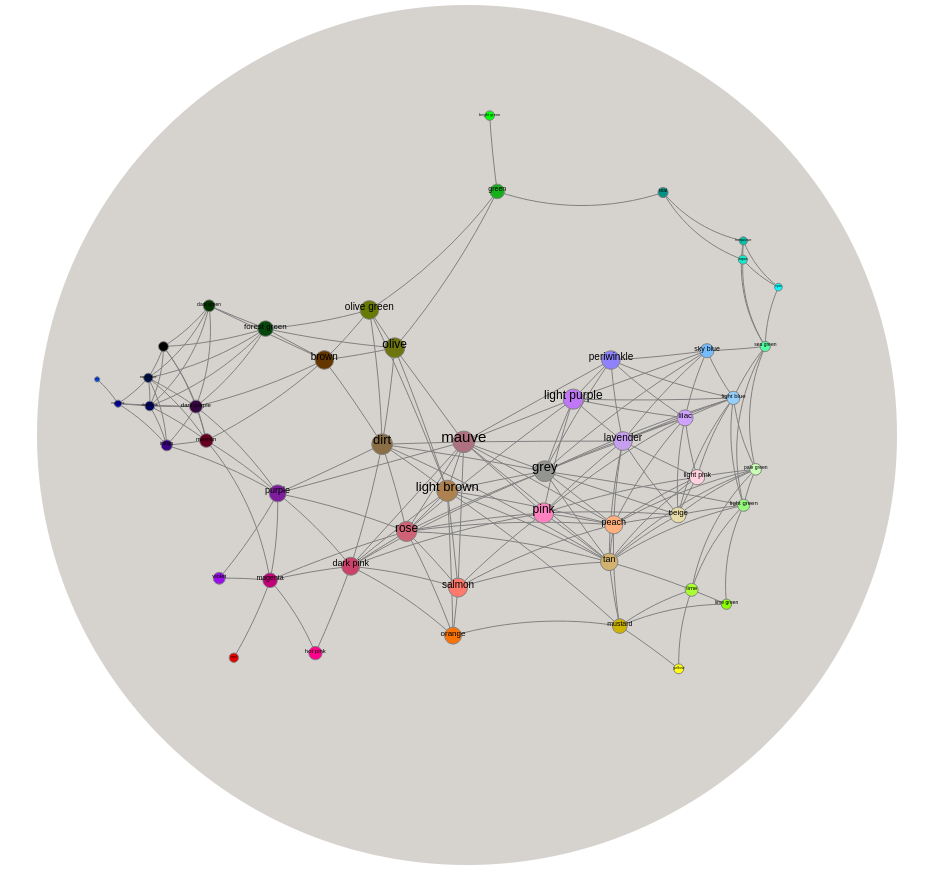}
        \includegraphics[width=.48\linewidth]{fixed_gmap.png}

        \includegraphics[width=.48\linewidth]{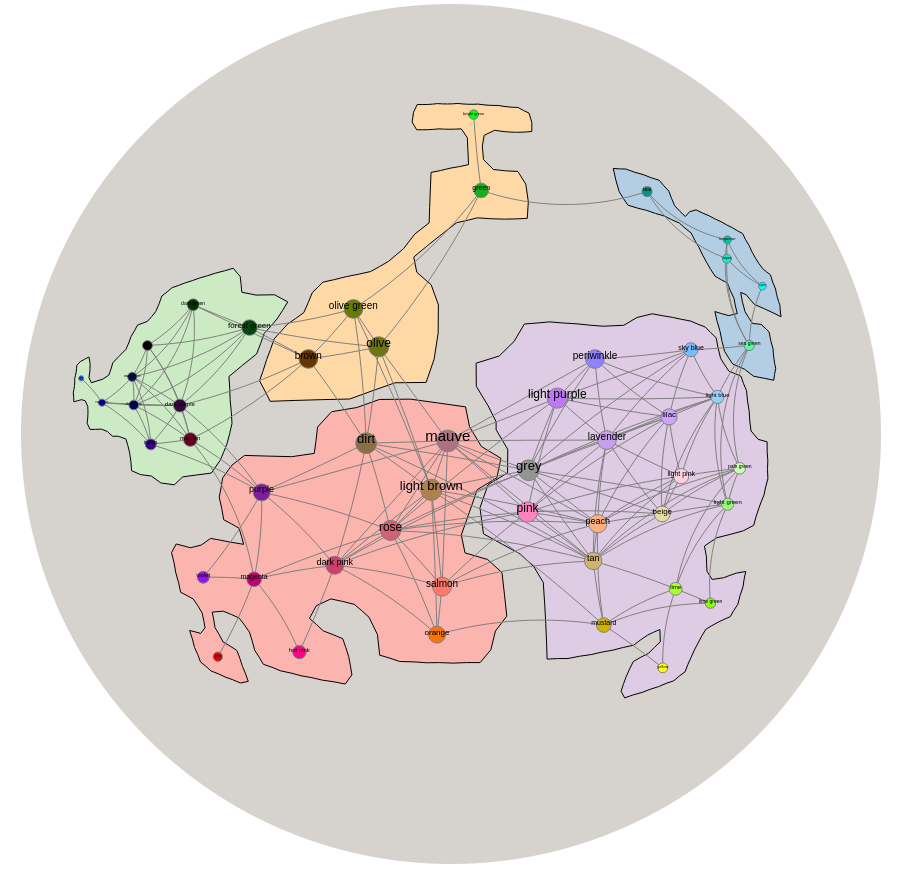}
        \includegraphics[width=.48\linewidth]{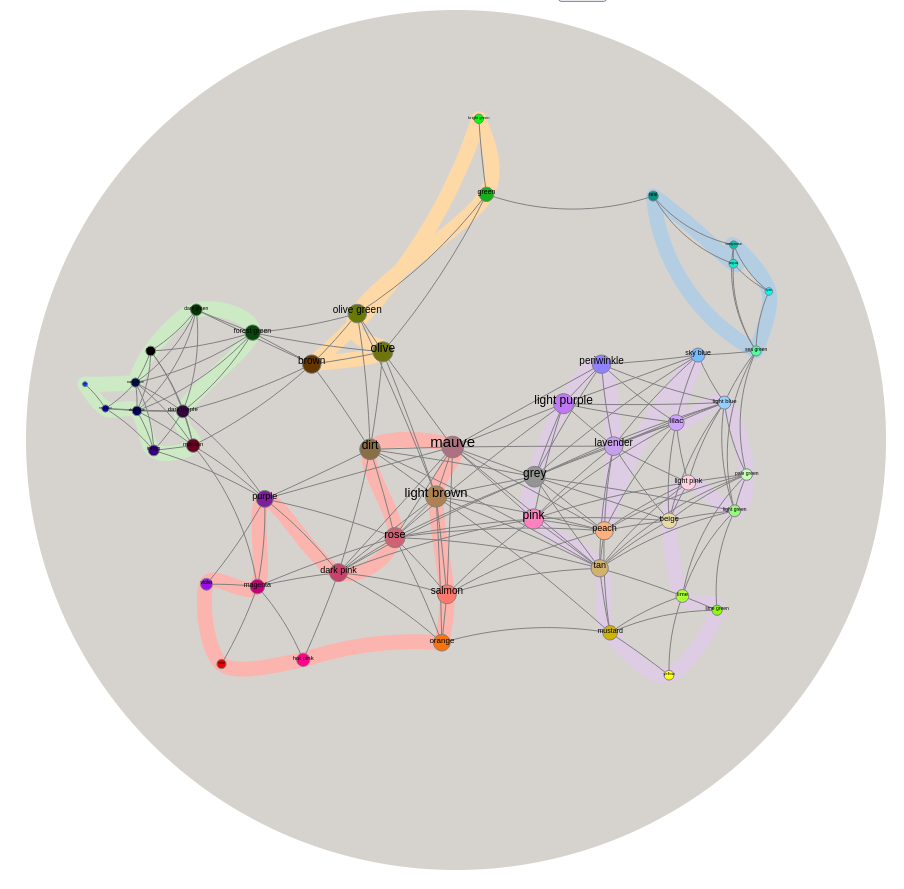}

    \caption[ ]
    {
    Different GMap drawing options for the same graph using inverse projection from Euclidean to hyperbolic space. 
    } 
    \label{fig:gmap-styles}
\end{figure}

The system relies on two different layout algorithms for computing a Euclidean layout: {\it sdfp} is a multi-level force-directed
algorithm~\cite{hu2005efficient} and {\em neato} is a implementation of the Kamada-Kawai algorithm~\cite{kamada1989algorithm}.
We show examples of the {\em colors graph} drawn as a node-link, GMap, BubbleSet, and LineSet diagram; see Fig.~\ref{fig:gmap-styles}.
This is a graph of the 38 most popular RGB colors, courtesy of xkcd\footnote{https://xkcd.com/color/rgb/}. 

We make use of a javascript library called Hyperbolic Canvas~\cite{hyperbolic-canvas}. It is a mathematical model of the Poincar\'e disk projection
of hyperbolic space that allows lines and shapes to be drawn using an HTML canvas. The projection-based pipeline below is based on the approach by Perry {\it et al.} for browser-based visualization of graphs on the sphere~\cite{perry2020drawing}.

\subsection{The Projection-based Pipeline}


Given a pre-computed 2-dimensional Euclidean layout, the projection-based method can be summarized as follows:
\begin{compactenum}
  \item Calculate geometric mean of the 2-d Euclidean layout
  \item Apply an inverse hyperbolic Lambert azimuthal projection centered on the geometric mean
  \item Project back into the Euclidean plane of the browser using the Poincar\'e projection (providing the look and feel of hyperbolic space). 
\end{compactenum}


\subsubsection{Hyperbolic Projections:}
It is well known that non-Euclidean spaces (such as spherical and hyperbolic spaces) can not be perfectly projected to the Euclidean plane. 
No matter what type of projection is used, something will get lost in the translation: distances are distorted, or region areas are distorted, or angles are distorted.
This problem is well studied in cartography in the context of projecting the sphere onto the 2-d Euclidean plane.

Knowing that a perfect embedding in the plane is impossible, useful maps can still be created by
 choosing which information to preserve. Three well-known projections of the sphere
are gnomonic, orthographic, and stereographic projections. The gnomonic projection preserves straight lines;
geodesics of the sphere are shown as straight lines in the projection. This is particularly useful in flight planning,
and is said to be the oldest map projection. The orthographic projection resembles the view of the Earth from space, and
preserves scale at the center of the projection, making it useful in visualization. Finally, the stereographic projection
preserves angles and has its roots in star charts used in sailing~\cite{snyderWorkingManual}. 

Hyperbolic surface is curved (negatively) just like spherical space is curved (positively), resulting in  similar problems when attempting to display it in the plane of a monitor or on a piece of paper. Just as
the sphere has many projections that serve different purposes, so there exists many hyperbolic
projections to the plane, although they are not as well studied. These projections can be thought of as analogous to their spherical counterparts
and can often be derived in an analogous way.
For instance, the Beltrami-Klein projection is analogous to the gnomonic projection of the sphere;
they both preserve geodesics as straight lines. Similarly, the Gans model of the hyperbolic plane
is analogous to the orthographic projection, in that they both have a point of perspective at infinity.
The Poincar\'e projection is a spherical analogue of the stereographic projection as they both preserve angles.

\begin{figure}[h]
  \centering
  \includegraphics[width=.32\linewidth]{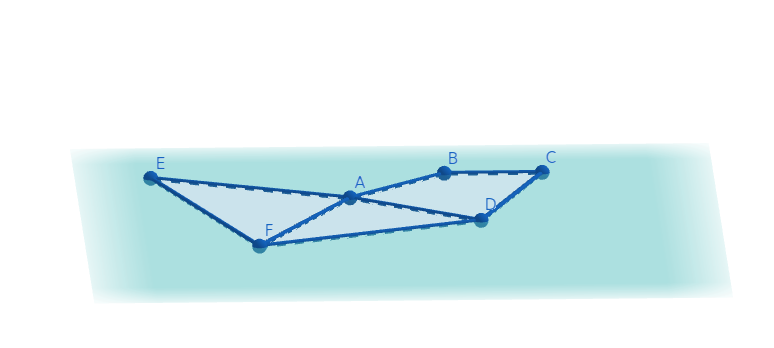}
  \includegraphics[width=.32\linewidth]{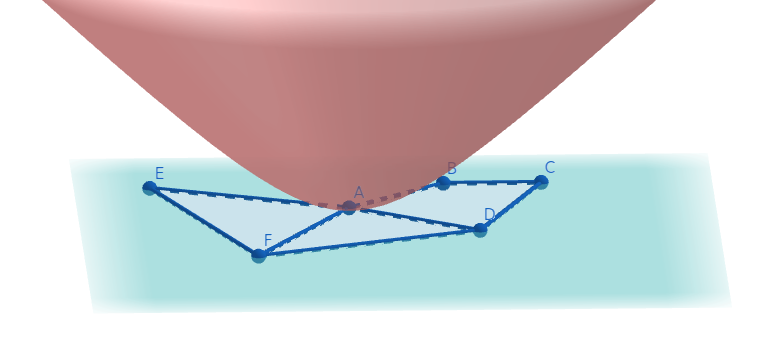}
  \includegraphics[width=.32\linewidth]{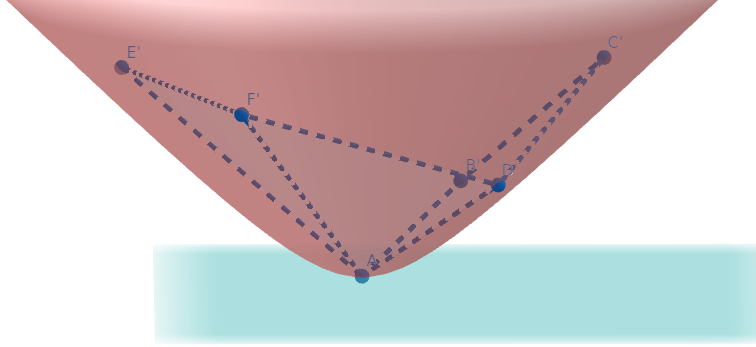}

\caption{Illustration of an inverse projection: wrapping a plane drawing on a hyperboloid.} 
\label{fig_lambert}
\end{figure}

\subsubsection{Hyperbolic Lambert Azimuthal Projection} Since we know we are projecting node-link and map diagrams, it seems
reasonable to choose to preserve areas. One way this can be accomplished is through a less common hyperbolic analogue to the Lambert azimuthal
projection, which has been called the hyperbolic Lambert azimuthal projection. This projection is equi-areal, so area is preserved. The hyperbolic
analogue can be derived in much the same way as the sphere.

Consider two disks: one in the 2-dimensional Euclidean space and the other in the 2-dimensional hyperbolic space. Denote the area of the Euclidean
disk of radius $r$ as $e(r)$ and the area of a hyperbolic disk of the same radius $h(r)$. We can then define the function $f(r)$ such that $e(r) = h(f(r))$. Assuming unit curvature, then
\[h(r) = 2\pi(cosh(r)-1)\]
\[e(r) = \pi r^2\]
\[f(r) = arccosh(\frac{1}{2}r^2+1)\]
where $arccosh$ is the inverse hyperbolic cosine. This maps the Euclidean plane to the hyperbolic plane and 
gives us the transformation $(r,\theta) \rightarrow (f(r),\theta)$, which preserves areas, but distorts angles and shapes.
The further away a shape is from the projection center  
the greater the distortion, 
 so centering about the geometric mean reduces this effect; see Fig.~\ref{fig_lambert}. 

\begin{figure}[ht]
  \centering
  \includegraphics[width=.32\linewidth]{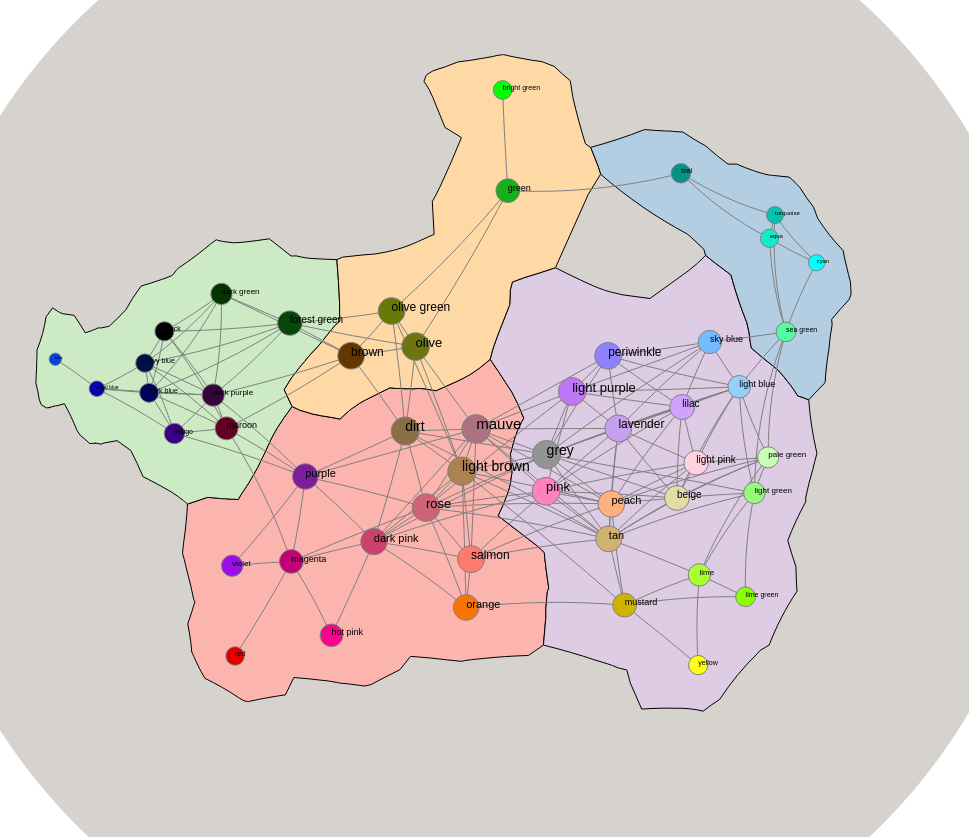}
  \includegraphics[width=.32\linewidth]{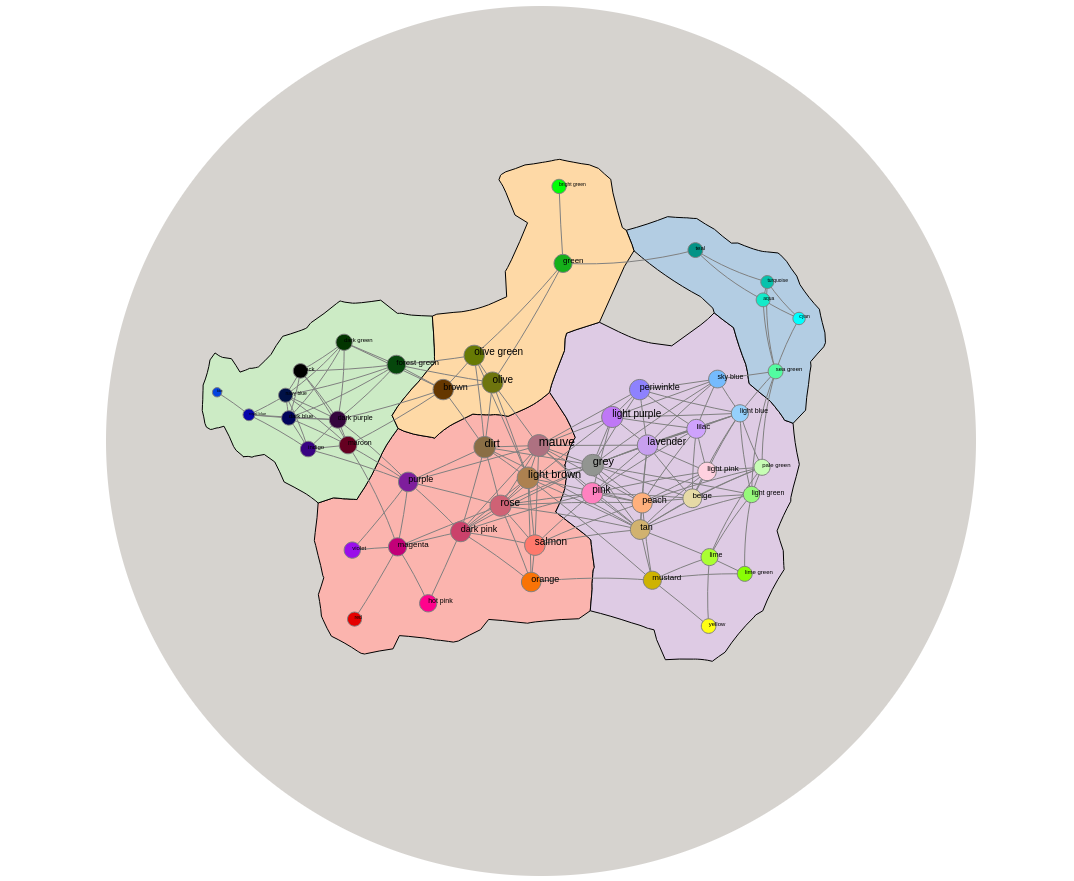}
  \includegraphics[width=.32\linewidth]{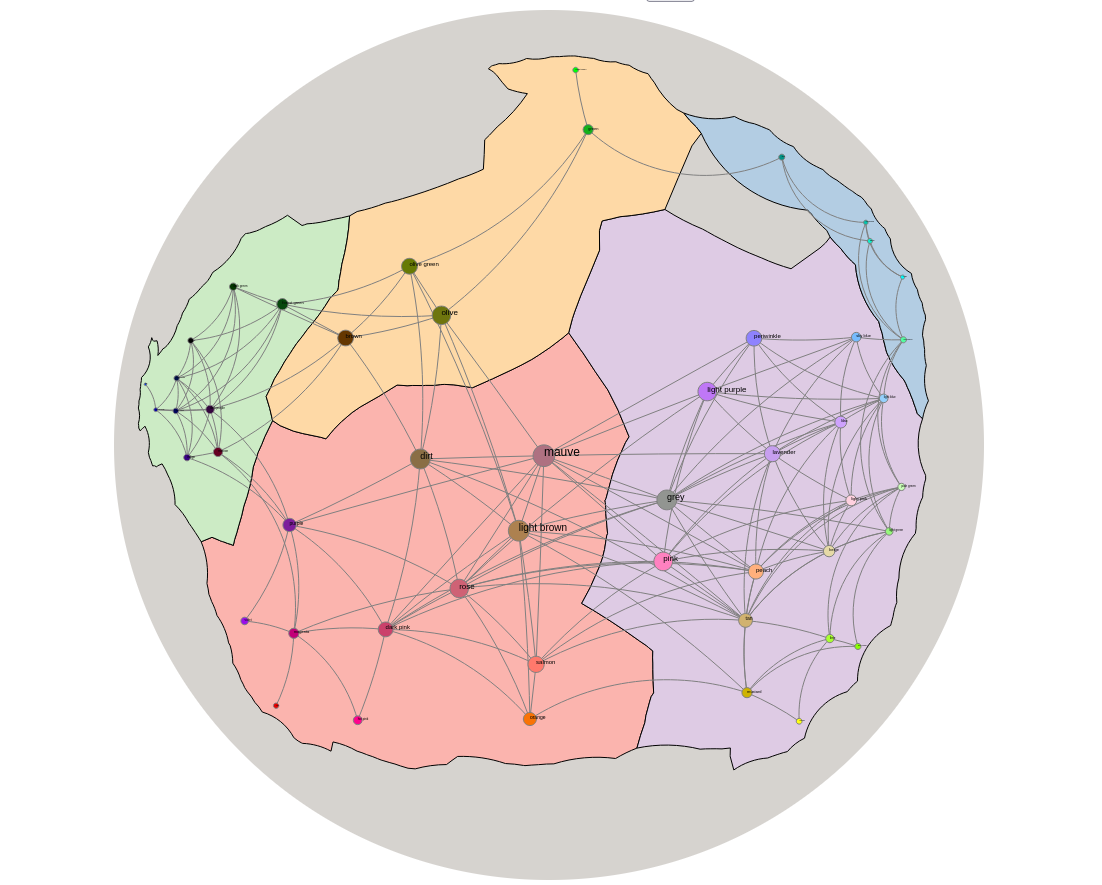}

\caption{An example of the default (center), increased zoom (left), and increased coverage (right) for the same graph.}
\label{fig:zoom-and-coverage}
\end{figure}

\subsubsection{Poincar\'e Projection}
Recall that the Poincar\'e projection of the hyperbolic plane is similar to the stereographic projection of the sphere, in that it preserves angles. 
The infinite hyperbolic plane is mapped to the inside of the unit disk with hyperbolic lines corresponding to either arcs of circles orthogonal to the boundary of the disk, or diameters of the disk if the line passes through the origin.
The Poincar\'e disk intrinsically provides the look and feel of hyperbolic space in the browser. The `focus+context' mentioned before is due to the Poincar\'e projection. A small area near the border of the disk represents a very large area in hyperbolic space, while the same size area near the center of the disk represents a small area of hyperbolic space. 
This can be seen mathematically in the  transformation that takes the hyperbolic plane to the Poincar\'e disk 
\[(r,\theta) -> (\frac{e^{r}-1}{e^{r}+1},\theta)\]

The exponentiation in the Poincar\'e transformation implies a practical limit on the hyperbolic radius of about 700, as larger values require dealing with large numbers and lead to numerical overflow. 

\subsection{Visualization Considerations}
In this section we discuss the interactive features, the parameters, and the task considerations.

\subsubsection{Navigating the Map:}
One of the main reasons for using map-like visualization for graphs is our familiarity with map interactions such as pan, zoom, click and drag. 
In the Poincar\'e disk, clicking and dragging brings new nodes and regions into focus, allowing the viewer to exploit the `focus+context' property of the projection. We accomplish this by making use of M\"{o}bius transformations.

A M\"{o}bius transformation is a complex function of the form $f(z) = \frac{az + b}{cz + d}$ where $z$ is a complex
variable and $ad-bc \neq 0$. M\"{o}bius transformations have many uses in complex analysis and geometry,
but one subgroup is especially useful for our purposes; the class of transformations that map the open unit disk to
itself. In particular the transformation
\[f(z) = \frac{z - z_0}{-\tilde{z_0}z+1}\]
takes $z_0$ to the origin and preserves the Poincar\'e projection of the hyperbolic plane, i.e., the transformation recenters the Poincar\'e projection at $z_0$.

\begin{figure}[ht]

  \centering
  \includegraphics[width=.48\linewidth]{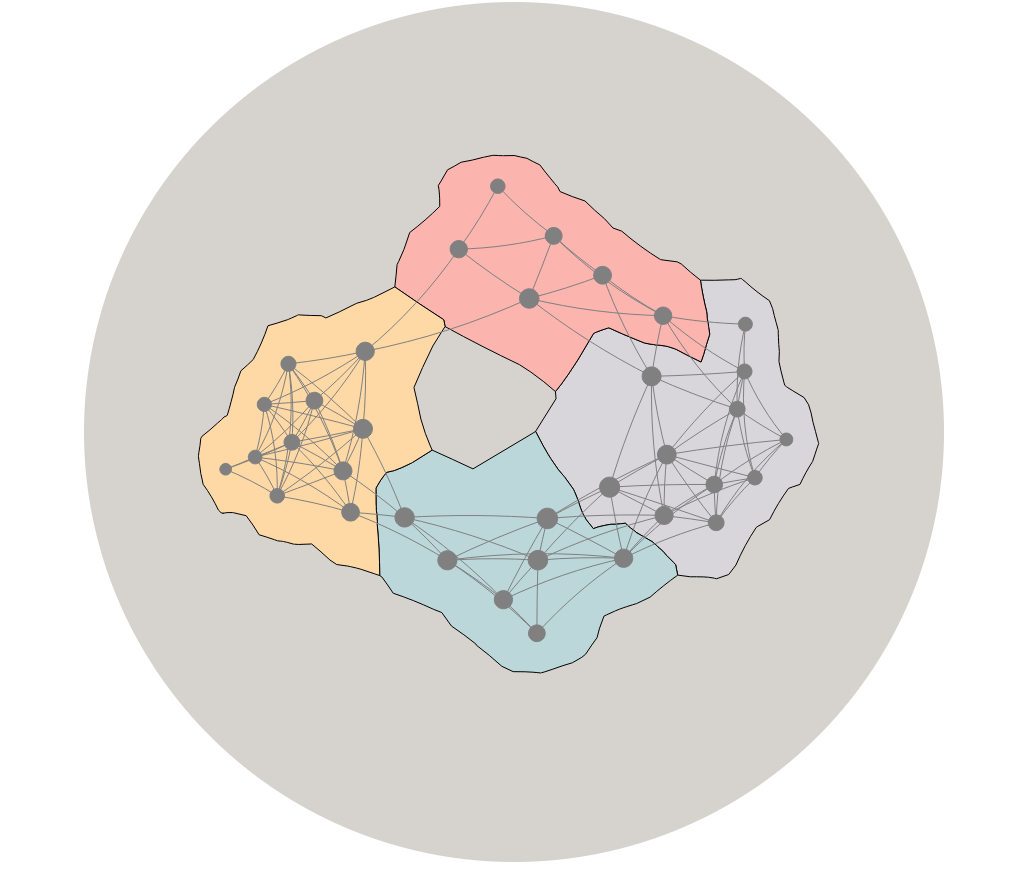}
  \includegraphics[width=.48\linewidth]{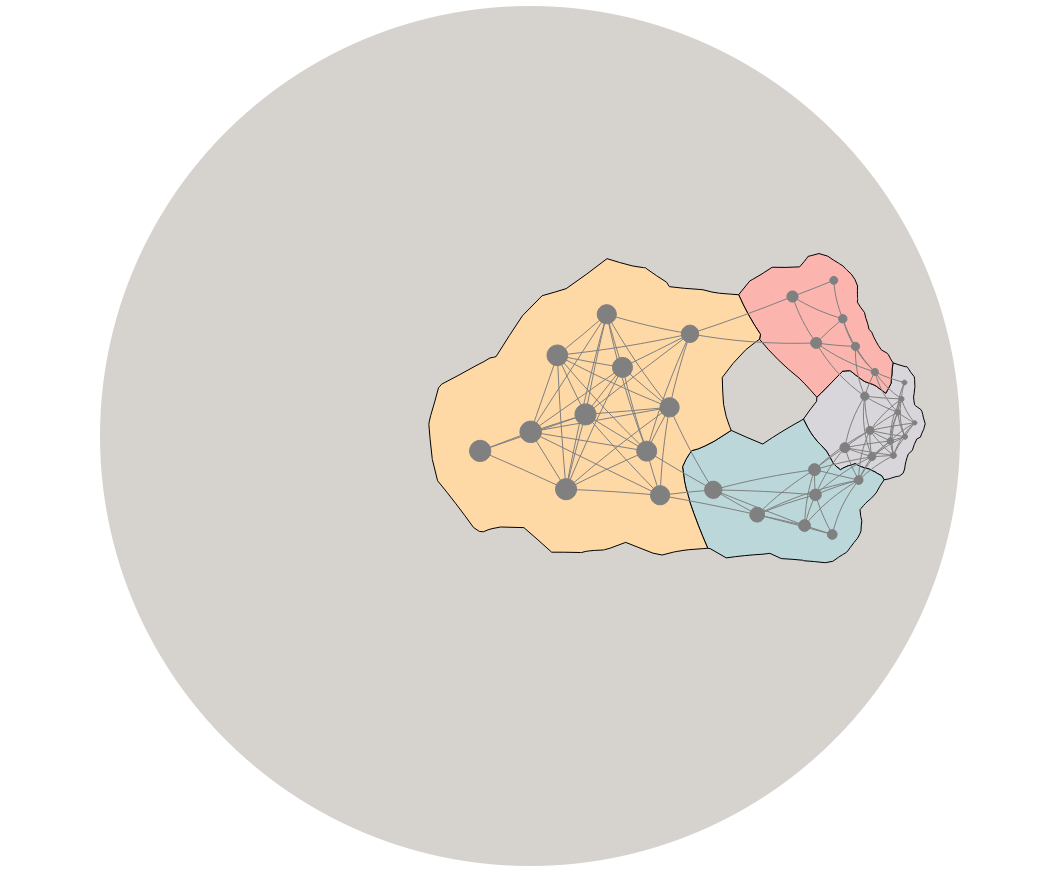}

\caption{The same graph centered about two different origins.}
\label{fig_moving}
\end{figure}

We can obtain transitions that look smooth to the human eye by repeatedly applying the above transformation
at a point some $\epsilon$ distance from the previous origin in the direction the mouse is
being dragged. 
Two still images centered at different points in a random graph are shown in Fig.~\ref{fig_moving}, but interacting with the actual visualization in GMap
better conveys the idea. Fig.~\ref{fig_music} additionally shows the difference between panning to the edge of a map in Euclidean space and hyperbolic space.

\subsubsection{Parameters}
There are a handful of settings in the parameter space, all of which can be hand-tuned within given bounds. 

{\bf Edge opacity: }controls the opacity of links. This is useful as the number of edges near the border in large graph can be so large that the layout is difficult to discern. An edge {\it opacity slider} helps mitigate this problem. 
The slider adjusts edge opacity from 0 (fully transparent) to 1 (fully opaque), which is the default setting.

{\bf Label size: }controls the relative size of labels in the view. The natural focus+context view of a graph in hyperbolic space is also reflected in the node label sizes, which decrease when farther away from the center. This process is automated by adjusting the font size of the labels in an inversely proportional fashion with respect to their Euclidean distance to the center of the disk. A {\it label size slider} allows the (central) font size to be adjusted from 0 (no labels) up to 40 px Arial. The initial setting is 15px, which ensures readability.

{\bf Zoom:} controls the size of the Poincar\'e disk in the browser window. Intuitively, the zoom slider brings the disk closer or further away from the point of perspective; see Fig~\ref{fig:zoom-and-coverage}. The slider begins at 100\% of the browser window size and the slider scales from 50\% to 150\% of the window size.

{\bf Coverage:} controls the total area the layout occupies in the Poincar\'e disk. 
As a consequence of Euclid's parallel postulate not holding in hyperbolic space, the hyperbolic plane is not invariant to scale~\cite{sala2018representation}. However, we can re-scale the
layout while it is still in the Euclidean plane, before we project it to $H^2$. By default we use an initial scaling factor proportional to the diameter (longest shortest path) of the graph. 
The scale can be adjusted from 50\% to 150\% of the default layout size.

\subsubsection{Tasks Considerations}
Brehmer and Munzner~\cite{brehmer2013multi} define an abstract task taxonomy based on a large body of related research. For domain specific tasks, Lee {\it et al.}~\cite{lee2006task} identify task abstractions for network data and Saket {\it et al.}~\cite{saket2014group,saket2014node} identify task abstractions for grouped node-link diagrams (e.g., map-like drawings). 

At a high level, we provide support for discover-type tasks by navigation. Node-link diagrams make adjacency easily apparent, and map-like drawings provide for straight-forward cluster identification. Panning aids in search tasks, such as location or exploration. A node/cluster can also be selected by double clicking, which smoothly recenters the layout around the selected node/cluster. 
Given the infinite space and  `focus+context' nature of the Poincar\'e projection, it is possible that a viewer may get lost. 
To alleviate this potential problem, we provide a {\it reset button}, that restores the layout to the original output of the underlying algorithm.

\begin{figure}[ht]

 \includegraphics[width=.48\linewidth]{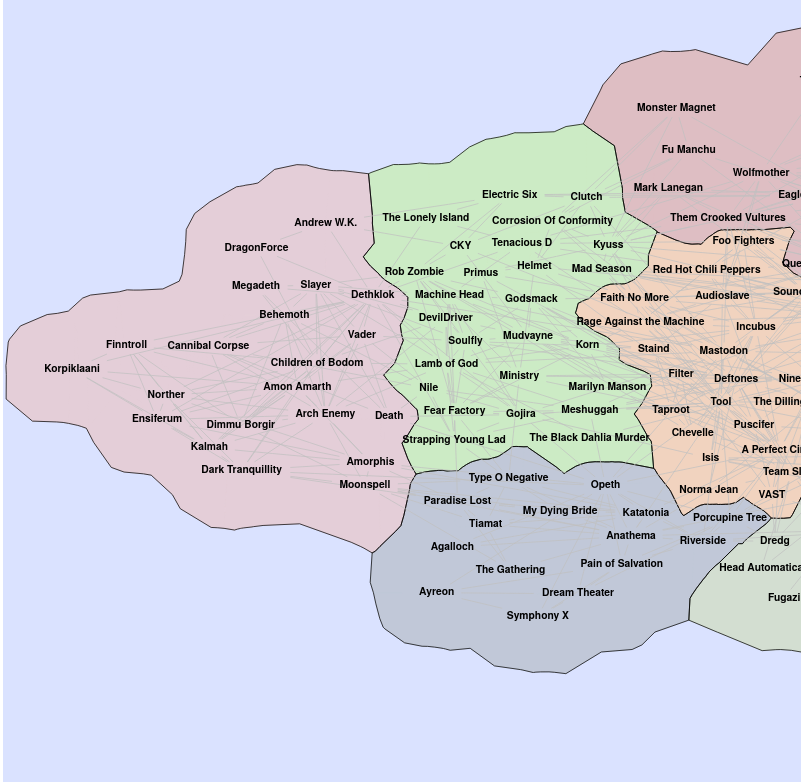}
  \includegraphics[width=.48\linewidth]{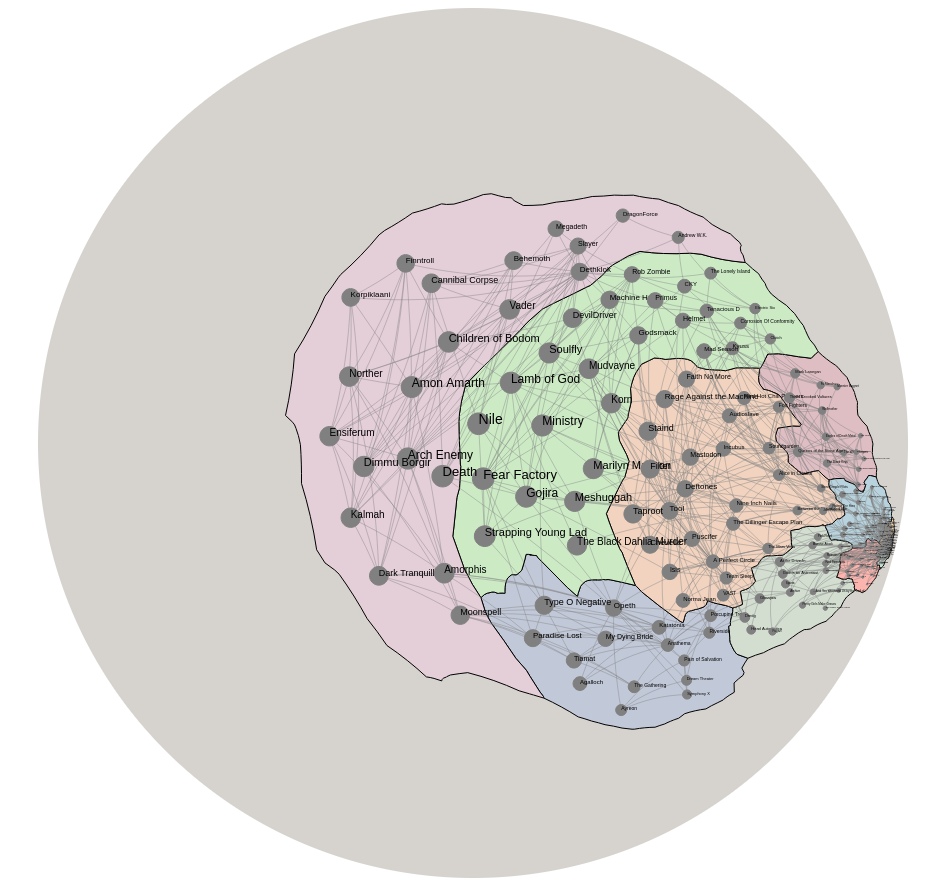}

\caption{A 2D Euclidean GMap instance of the MusicLand graph (left) and its hyperbolic realization (right).} 
\label{fig_music}
\end{figure}


\section{Force-directed Method}
Our projection-based hyperbolic visualization method uses a precomputed 2-dimensional Euclidean layout, but it uses hyperbolic space just for the visualization and `focus+context' effect, rather than for the actual graph embedding. 
Properly embedding the graph in hyperbolic space would allow us to take advantage of the underlying hyperbolic geometry. Algorithms for directly embedding special classes of graphs in hyperbolic space, such as trees and hierarchies, can better take advantage of the properties of the space and obtain better embeddings than via projections. It is also possible to modify the standard force-directed algorithm for operation in  Riemannian geometries (such as hyperbolic and spherical) by taking advantage of the locally Euclidean properties of such spaces~\cite{kobourov2005non}.
The implementation, which provides visualization in the browser, and is made available in a browser based system through GitHub.

The idea is to compute a tangent plane at each vertex embedded in the non-Euclidean Riemannian space, mapping every other vertex to that plane, performing
a step of a force-directed algorithm in the plane, and projecting back the resulting node position changes to the Riemannian space.
While conceptually simple, this method allows the graph to make use of the properties of the corresponding non-Euclidean geometry.

For instance, on the sphere, layout methods that correctly make use of the geometry allow 3D polytopes to wrap ‘around’ the sphere. Thus, compared to the plane, a more accurate realization of their structure is possible; see method two of~\cite{perry2020drawing}.


We apply this idea to the Kamada-Kawai type of force-directed graph layout algorithm, for its conceptual simplicity and
its desirable property of capturing graph structure (e.g., graph distances between pairs of nodes) in the embedding (e.g., realized distances between pairs of nodes in the non-Euclidean space).  
Specifically, we compute the graph theoretic distances between all pairs of nodes and these define desired distances in the layout. Spring forces, proportional to the squared Euclidean distance between nodes in
the layout, are used to gradually improve a given initial layout to one in which realized distances match the graph theoretic distances~\cite{kamada1989algorithm}.
 Formally, there is an attractive or repulsive force (similar to  stress) defined for any pair of edges based on the difference between the graph theoretical distance and the realized distance in the current embedding. Specifically, the total energy of the system is modeled as:
 \[E = \sum_{i=1}^{n-1}\sum_{j=i+1}^{n} \frac{1}{2}k_{ij}(|p_i - p_j| - d_{ij})^2\]
 where given a pair of nodes $i$ and $j$, $d_{ij}$ is the graph theoretic distance between them, $|p_i - p_j|$ is the current realized  distance in the embedding between them, and $k_{ij}$ is the strength of the spring forces between them.
 The layout is obtained by reducing the energy of the system via gradient descent. 

\begin{figure}[ht]
\centering
\includegraphics[width=.45\linewidth]{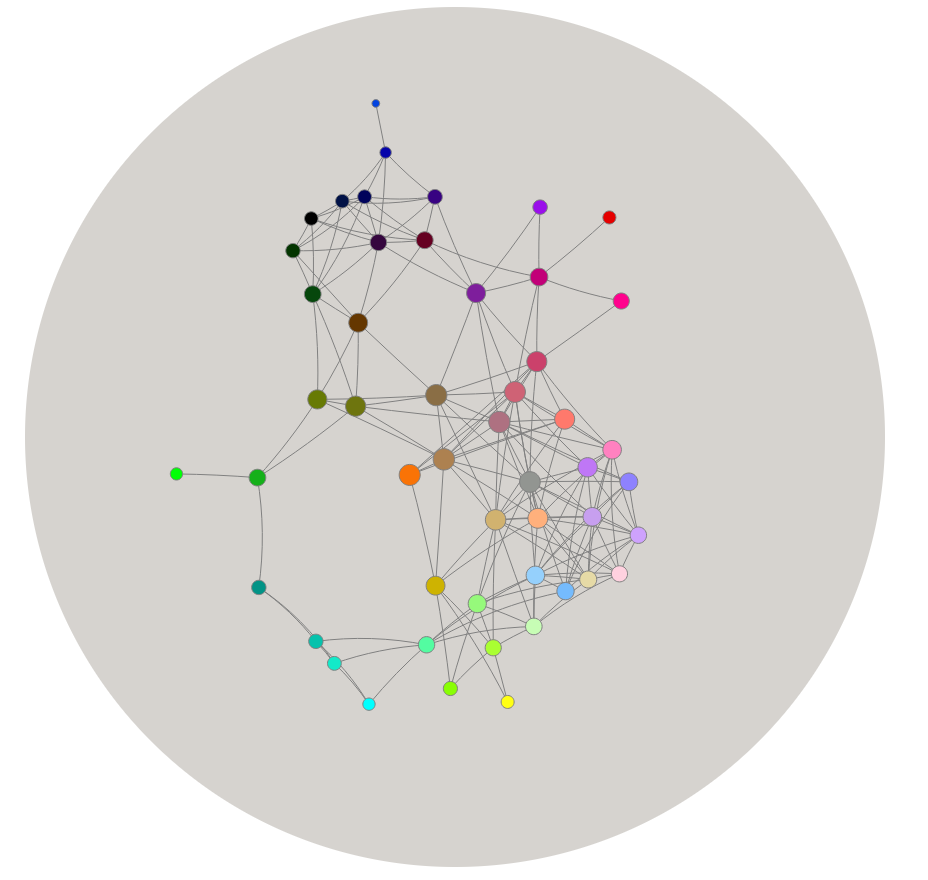}
\includegraphics[width=.53\linewidth]{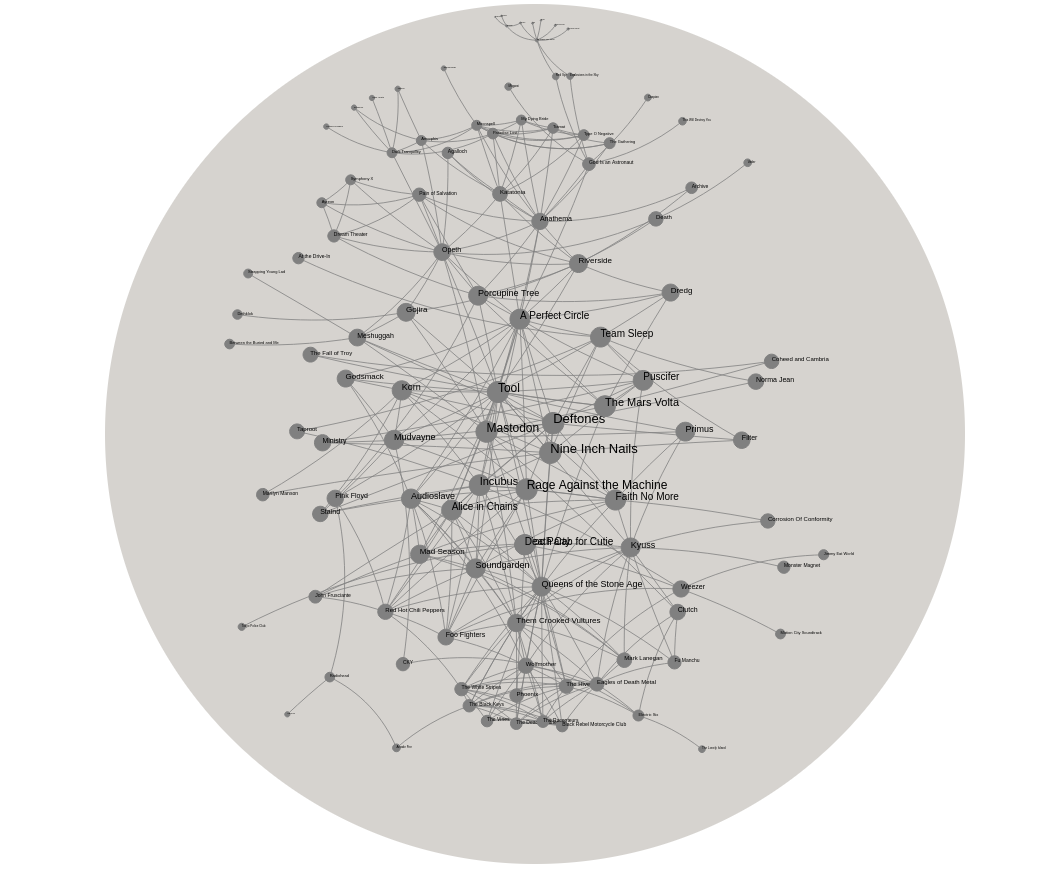}
%
\caption{Force-directed colors (left) and MusicLand (right) graphs.}
\label{fig_musicLandComparison}
\end{figure}




\subsection{Tangent Plane}
In order to compute a tangent plane at some node $x$ in $H^2$, we need to set the distance between $x$ and every other node in the plane to the hyperbolic distance between them, and ensure the angle between the nodes stay the same~\cite{kobourov2005non}. 
The Poincar\'e disk preserves angles, so we only need to map hyperbolic to Euclidean distances. In the Poincar\'e model, hyperbolic distance is 
simplest to compute from the origin, so we first apply the M\"{o}bius transformation that takes $x$ to the origin.
The distance between $x$ and any node $y$ is 
\[d_h(x,y) = ln(\frac{1+|y|}{1-|y|}) = 2arctanh\: |y|\]
where arctanh is the inverse hyperbolic tangent. 
Then, let $x$ be the center of the disk and for every node $y$, let the transformation $(|y|,\theta) \rightarrow (d_h(x,y),\theta)$ 
be its location in the tangent Euclidean plane, using polar coordinates. 

Once the tangent plane is computed and a step of the force-directed algorithm has completed, the central node must be placed back into hyperbolic space. 
This is accomplished through an inverse of the above equations. Let $y'$ be the new location of the moved node. We apply the transformation 
\[(|y'|,\theta)->(tanh\:\frac{|y'|}{2},\theta)\]
The following M\"{o}bius transformation takes the disk back to its former origin, $z_0$
\[f^{-1}(y) = \frac{-y -z_o}{-\tilde{z_o}y - 1}\]

\subsection{Precision}
While hyperbolic geometry poses many interesting challenges for graph drawing, the
most notable we encountered was the issue of precision. It is well known that floating point
numbers are not arbitrarily precise and that this can cause problems when the number of significant
bits needed is large. This effect is pronounced on the Poincar\'e disk, as the number of
bits needed to accurately reflect the hyperbolic position increases exponentially as one approaches
the border of the disk.
%
We choose to trade
accuracy for stability: using the Euclidean coordinates of the unit disk, if a node is pushed to within 0.001 of the border, the node is `pulled back' to avoid errors from `ideal' points of magnitude $\geq$ 1.  
This effectively creates a bounding region defined by a circle of (Euclidean) radius 0.999 from the center of the Poincar\'e disk. This precision could be increased to allow for larger nodes extremely far away from the center, or decreased to keep the visual distortion of the drawing small.

\subsection{Maps}
Once we have computed a layout for a node-link diagram, we can compute a map-like representation for it by projecting it to the plane and running the existing GMap/BubbleSets/MapSets/LineSets algorithm to obtain the needed groups and polygons. 

It should be possible to compute the map-like representations directly in hyperbolic space. For example, the cluster regions (polygons) for GMaps are computed using Voronoi diagrams and it has been shown that Voronoi diagrams for 2-dimensional points generalize to hyperbolic space and can be computed in $O(n \log n)$ time~\cite{nielsen2010hyperbolic}. Similarly, LineSets requires Bezier curves between nodes in a cluster, which should also be computable in hyperbolic space.




\section{Multidimensional Scaling in $H^2$}
In the force-directed approach, we compute the graph embedding with the help of many tangent plane computations, so that we can use the standard force computations in Euclidean space. 
Here, we consider a simple  embedding: hyperbolic multidimensional scaling (H-MDS).

Recall that metric multidimensional scaling is a dimensionality reduction technique that attempts to preserve relationships between $n$ data points by finding a set of $n$ points in the target space whose distances match observed distances. 
MDS can be naturally formulated as a graph drawing problem by computing the pairwise distances through an all-pairs-shortest-paths computation. Metric MDS is then typically solved by minimizing an objective function. The most common function used is known as stress
\begin{equation}
\label{eq:stress_def}
\mathit{Stress} = \sum_{i<j}w_{ij}(\Vert X_i - X_j \Vert - d_{ij})^2
\end{equation}
where $d_{ij}$ is the given
distance between two nodes in the graph,
$\Vert X_i - X_j \Vert$ is the distance between them in the target space, and $w_{ij}$ is a normalization factor (typically 1 if the given distances are of the same order, or $d_{ij}^{-2}$ if the given distances include both very large and very small distances). 
The stress function is non-convex and classic optimization techniques are not guaranteed to find the global optimum. However, existing techniques, such as gradient descent, stochastic gradient descent and stress majorization, achieve sufficiently good results in practice.

Early approaches for H-MDS suggest using gradient descent~\cite{walter2004hmds}. However, gradient descent is too slow in practice even for relatively small graph sizes. We adapt stochastic gradient descent (SGD) to provide reasonable runtimes for the browser.

The SGD algorithm considers random pairs of nodes, calculates the minimum distance the pair needs to be moved to realize its observed distance $d_{ij}$, then steps along this direction by a distance proportional to the learning rate. 
As we aim to find an embedding in hyperbolic space, in order to evaluate the stress function and perform gradient descent, we need to choose an appropriate coordinate system. 
%
We use a coordinate system known as Lobachevsky coordinates in order to solve the H-MDS problem via SGD.
Lobachevsky coordinates are defined as a pair of real numbers $(x,y),$ where for a given point, $x$ is its distance along a geodesic horizontal axis and $y$ is its perpendicular distance to that axis. 
Lobachevsky coordinates for hyperbolic space are analogous to Cartesian coordinates for the Euclidean space. 
For example any pair of real numbers represents a point in hyperbolic space and any point in hyperbolic space can be represented by a pair of real numbers.

%

\subsection{Parameters}

The SGD algorithm depends on several parameters and tuning these parameters can have non-trivial impact. Here we discuss these parameters and how we set the default values.

\subsubsection{Randomization}
\label{sec:shuffle}

Computing the stress function for a given embedding requires $O(|V|^2)$ time by definition; see \eqref{eq:stress_def}.
Thus the gradient computation also requires quadratic runtime per iteration.
SGD allows for better runtimes in practice,  using constant-time computations at each step by only calculating the gradient of a specific pair at a time, although this pairwise gradient calculation needs to be performed many more times. 
The classic SGD method samples the original data with replacement~\cite{robbins1951stochastic}. In our case, this would mean choosing two nodes at random every step until complete. On the opposite end, a method known as random reshuffling enumerates all possible data subsets (in our case all pairs) and shuffles this list. This ensures that while the order of pairs moved is random, each pair is guaranteed to be moved once in a fixed number of steps. Under certain conditions, random reshuffling outperforms and converges faster than classical replacement~\cite{gurbuzbalaban2021random} and has been shown to work well for Euclidean SGD~\cite{DBLP:journals/tvcg/ZhengPG19}. A third method known as index shuffling randomizes the indices in place, and pairs are chosen from this new ordering.

\begin{figure}
    \centering

        \includegraphics[width=0.47\linewidth]{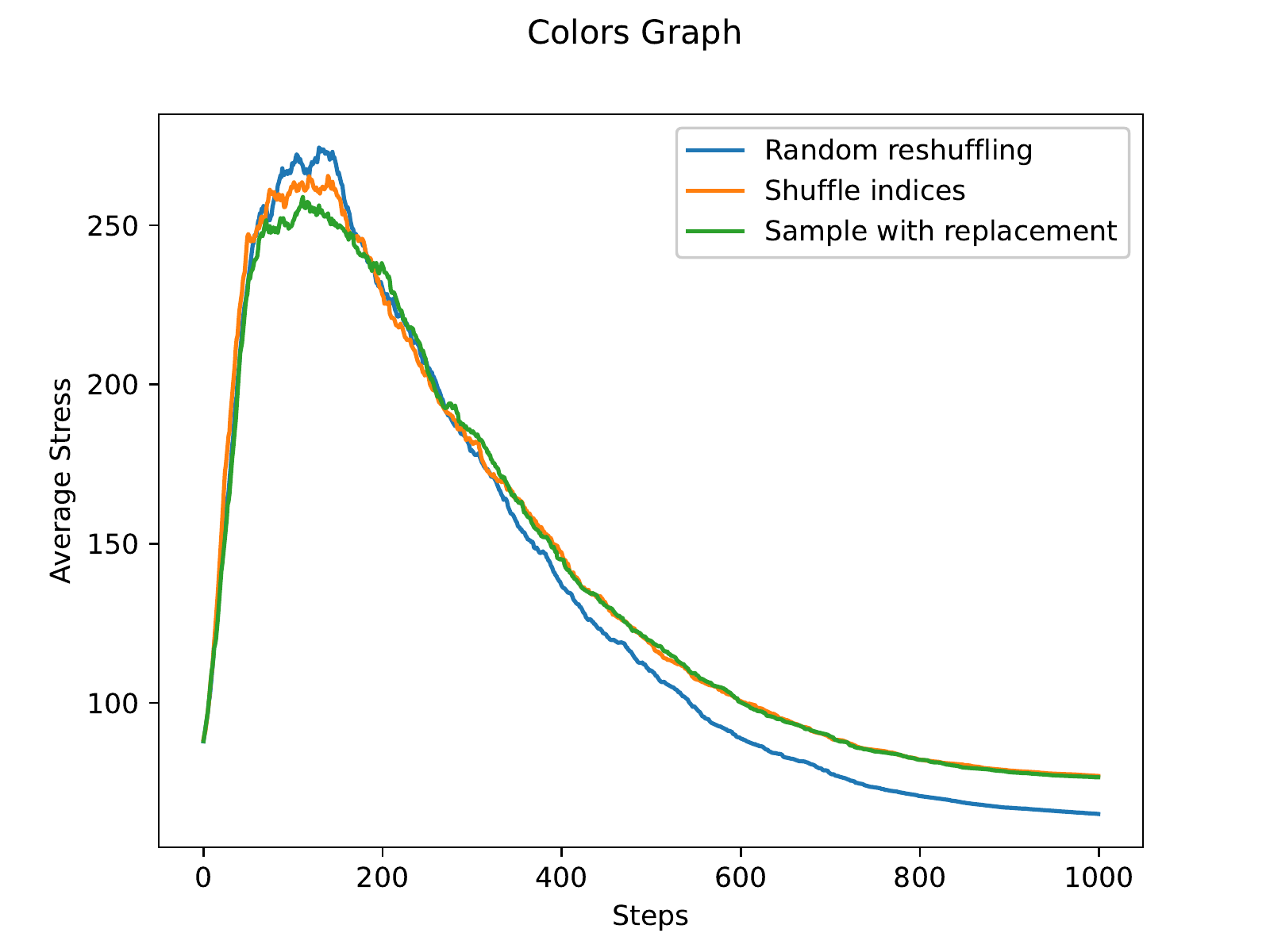}
        \includegraphics[width=0.47\linewidth]{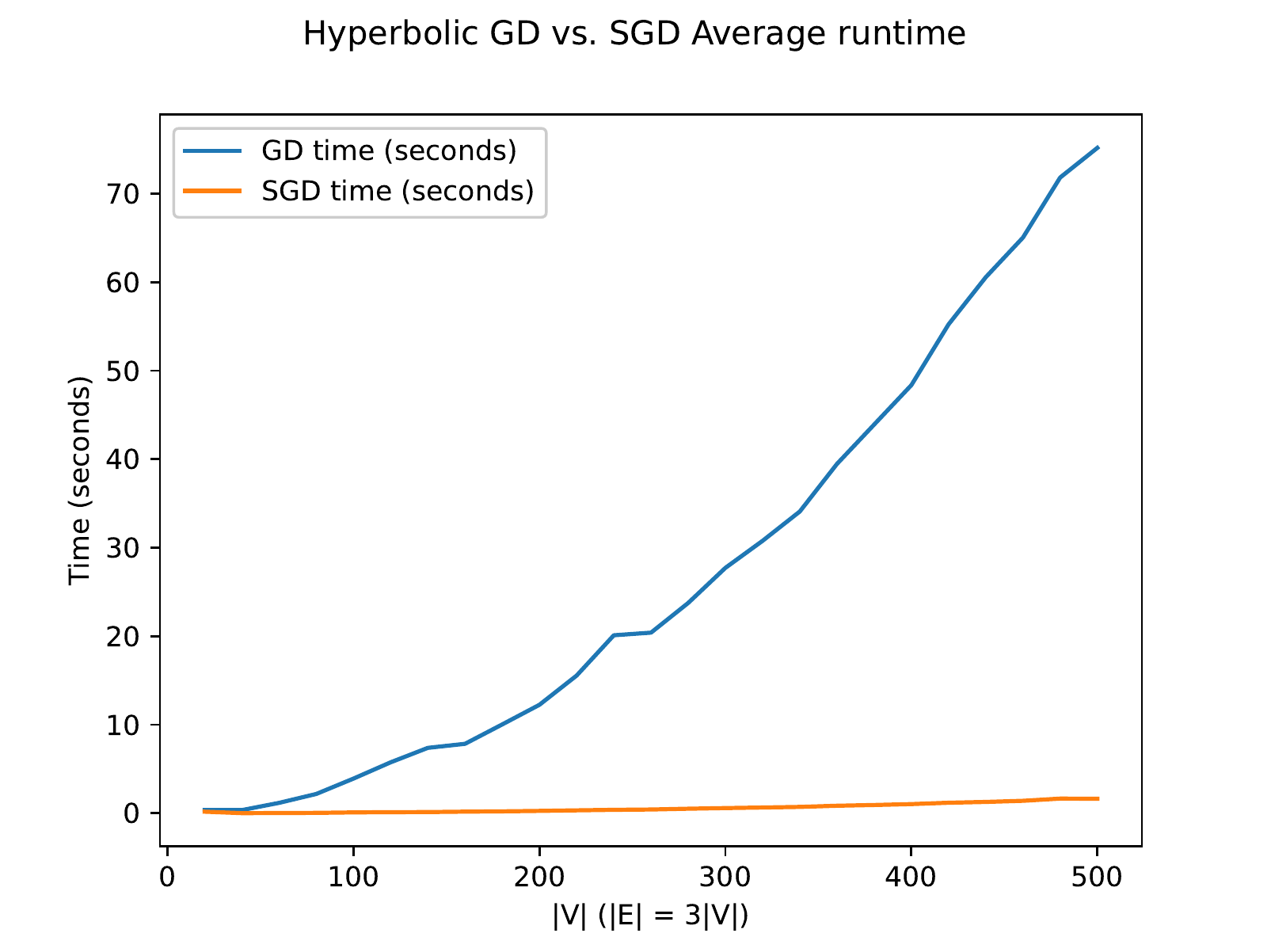}

    \caption{ Effect of randomization techniques (left) described in \ref{sec:shuffle}, showing average stress over 30 runs at each step on the Colors graph and average runtime (right) of HMDS using GD and SGD over 30 runs (pre-processing omitted). Similar results were found on other graphs.} 
    \label{fig:random-effects}
\end{figure}

We investigate these three randomization methods in the context of H-MDS: classical sampling with replacement, shuffling of indices, and random reshuffling. 
We demonstrate that random reshuffling tends to reach the lowest stress values; see Fig.~\ref{fig:random-effects}.
We use random reshuffling and define an {\it iteration} as a full pass through all pairs and show, in Section~\ref{sec:iterations}, that we only need a constant number of iterations.

\subsubsection{Initialization}
As the stress function \eqref{eq:stress_def} is non-convex, there are no convergence guarantees for gradient descent or for SGD. 

Recent work has shown that `smart initialization' is not necessary for SGD in Euclidean space, as the algorithm is  consistent regardless of initial embedding~\cite{DBLP:conf/gd/BorsigBP20}. To see if this holds true for hyperbolic space, we performed a small-scale analysis on a selection of graphs (chosen from the sparse matrix collection~\cite{DBLP:journals/toms/DavisH11}) and compared our smart initialization to random initialization. 

Knowing the Euclidean algorithm is good at escaping local minima, we run Euclidean SGD on the graph for 5 iterations, then project this layout into hyperbolic space to obtain our smart initialization. Random initialization is obtained by placing each node uniformly at random in a circle of hyperbolic radius 1. While we initially saw small improvements, there was no statistically significant benefit of smart initialization over using a random initialization, confirming the results of~\cite{DBLP:conf/gd/BorsigBP20}.

\subsubsection{Learning Rate}
Another important parameter for SGD is the learning rate.
At each iteration of gradient descent, we move the value being optimized along the steepest direction of the gradient by a size proportional to the learning rate, $\eta$. 
If the learning rate is very small, the algorithm might take too long to converge; if the learning rate is too large, the algorithm might not converge. Thus, the proper choice of learning rate is crucial for both the accuracy and the speed of the algorithm.

Generally, it is good for $\eta$ to be large for the initial steps to move the system quickly to a lower energy configuration, but $\eta$ should tend toward zero as the number of iterations increases so that
 the algorithm converges. 
Computing a good $\eta$ is a research topic all on its own and is important for SGD's effectiveness~\cite{ruder16,DBLP:conf/gd/BorsigBP20}.
We upper bound the product $\eta w_{ij} \leq 1$ as in~\cite{DBLP:journals/tvcg/ZhengPG19}.
This allows us to use a larger initial rate to `jump' out of 
bad neighborhoods and possible local optima, 
but still converge 
as $\eta$ goes to $0$. 
We set a maximum and a minimum learning rate, a function $s(t)$ that outputs a 
learning rate
$\eta$ at time step $t_i$. $s(t_0) = \eta_{max} = d_{max}^2$ and $s(t_{max} = \eta_{min} = \epsilon d_{min}^2$ where $d_{max}$ and $d_{min}$ correspond to the longest and shortest shortest paths of the input graph, respectively. 

Euclidean SGD works particularly well with an exponential decay learning rate~\cite{DBLP:journals/tvcg/ZhengPG19}. 
To test if hyperbolic SGD behaves the same way, we compare this exponetial decay learning rate with two additional schedules: $\Theta(1/t)$ and $\Theta(1/\sqrt{t})$ schedules. 
We define the exponential schedule according to~\cite{DBLP:journals/tvcg/ZhengPG19} using $\eta_{max}e^{-bt}$, the traditional $\Theta(1/t)$ as $\frac{a}{1+bt}$ and the $\Theta(1/\sqrt{t})$ schedule as $\frac{a}{\sqrt{1+bt}}$. 
We set $a = d_{min}^2$ and $b = -(t_{max})log\frac{\eta_{min}}{\eta_{max}}$.

\begin{figure}
    \centering
  \includegraphics[width=.48\linewidth]{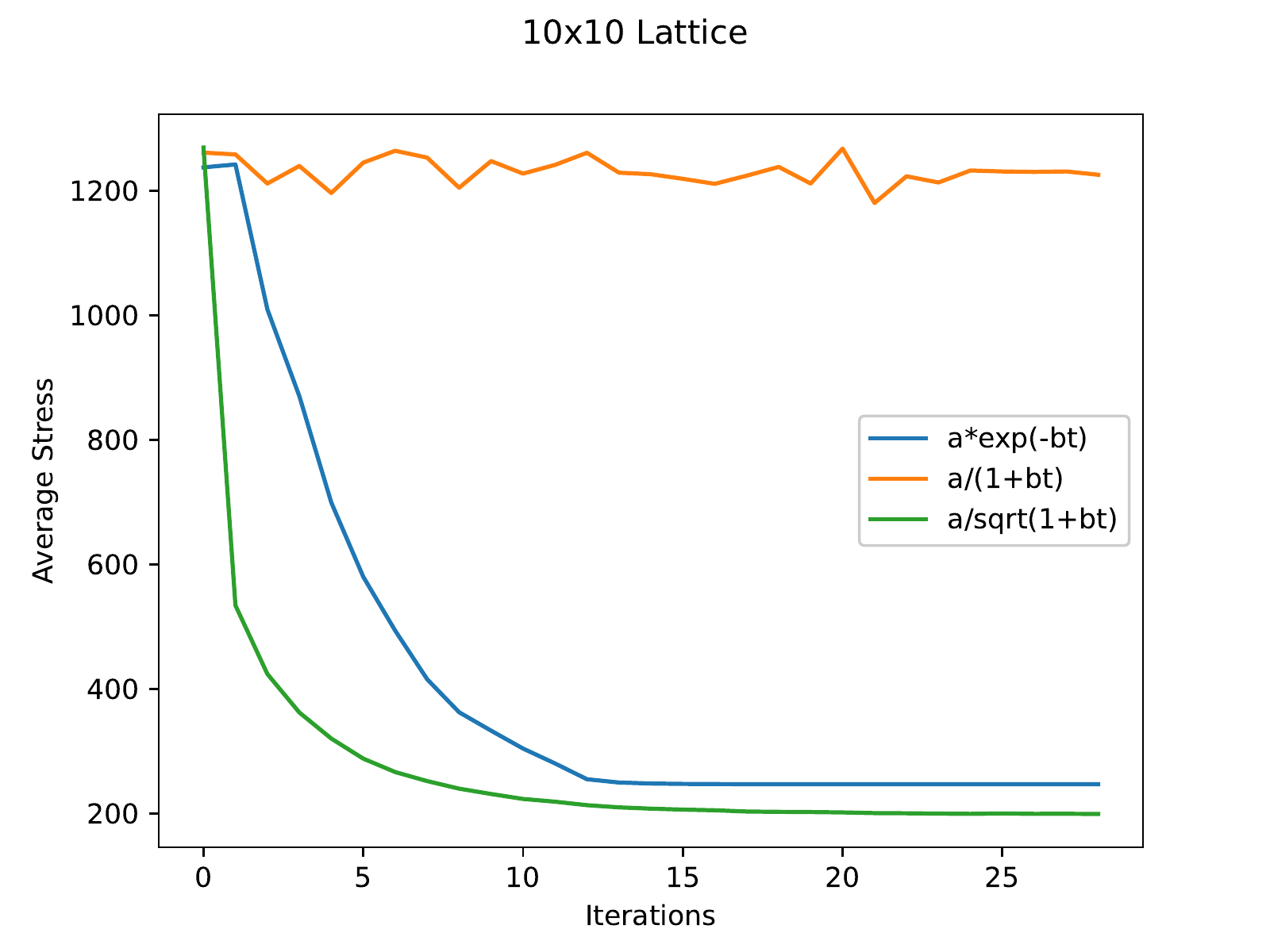}
  \includegraphics[width=.48\linewidth]{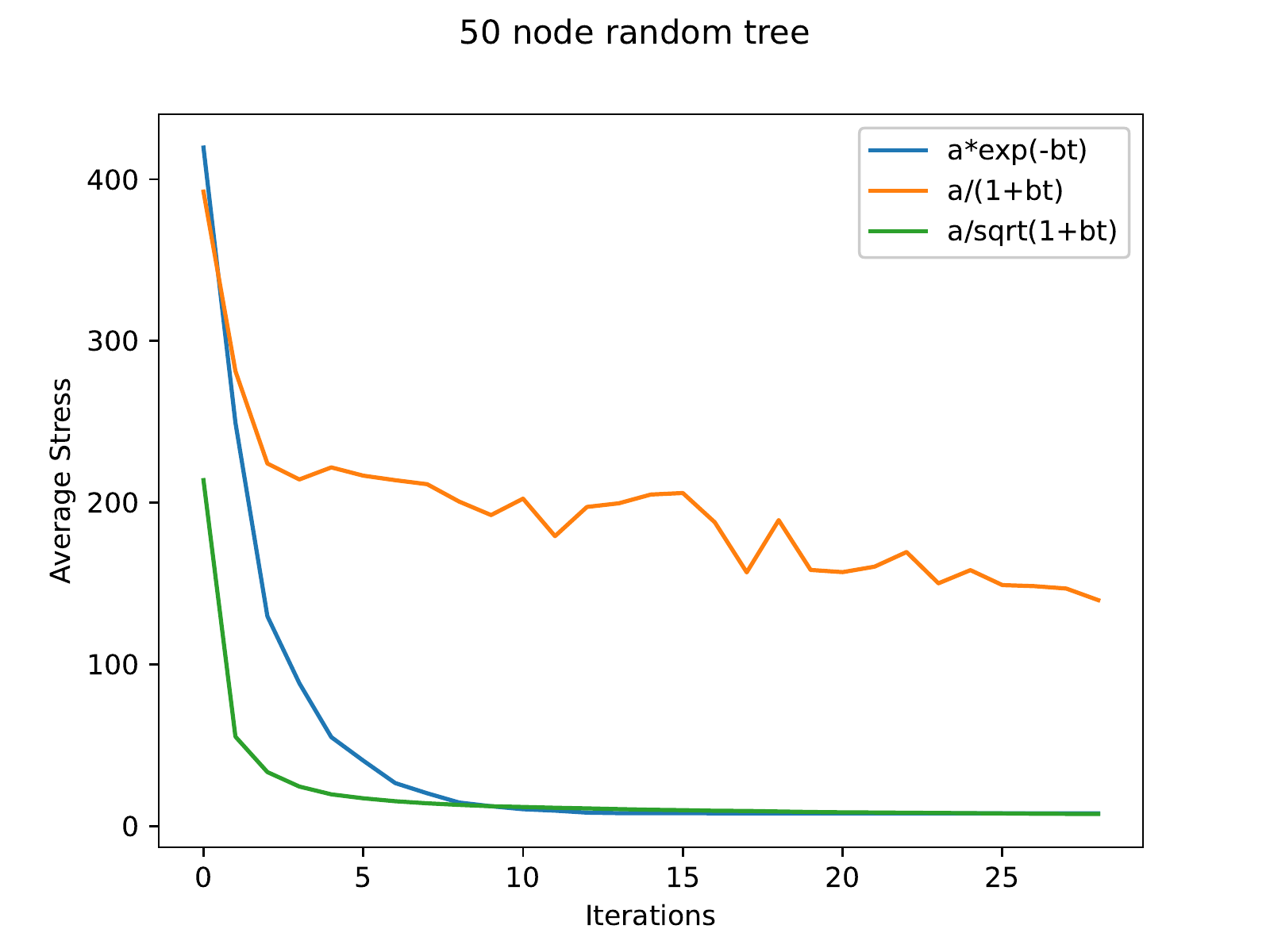}
  
  \includegraphics[width=.48\linewidth]{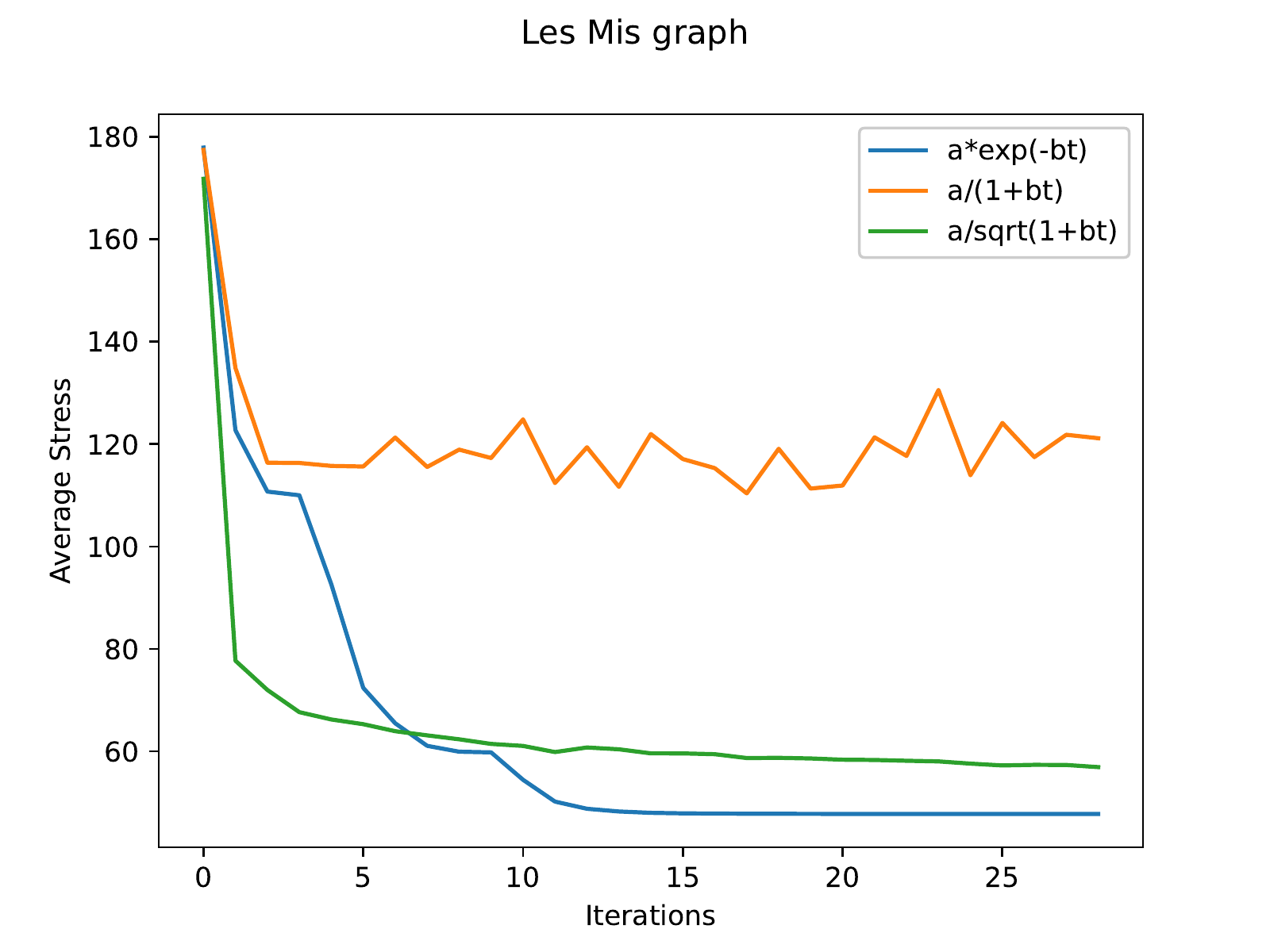}
  \includegraphics[width=.48\linewidth]{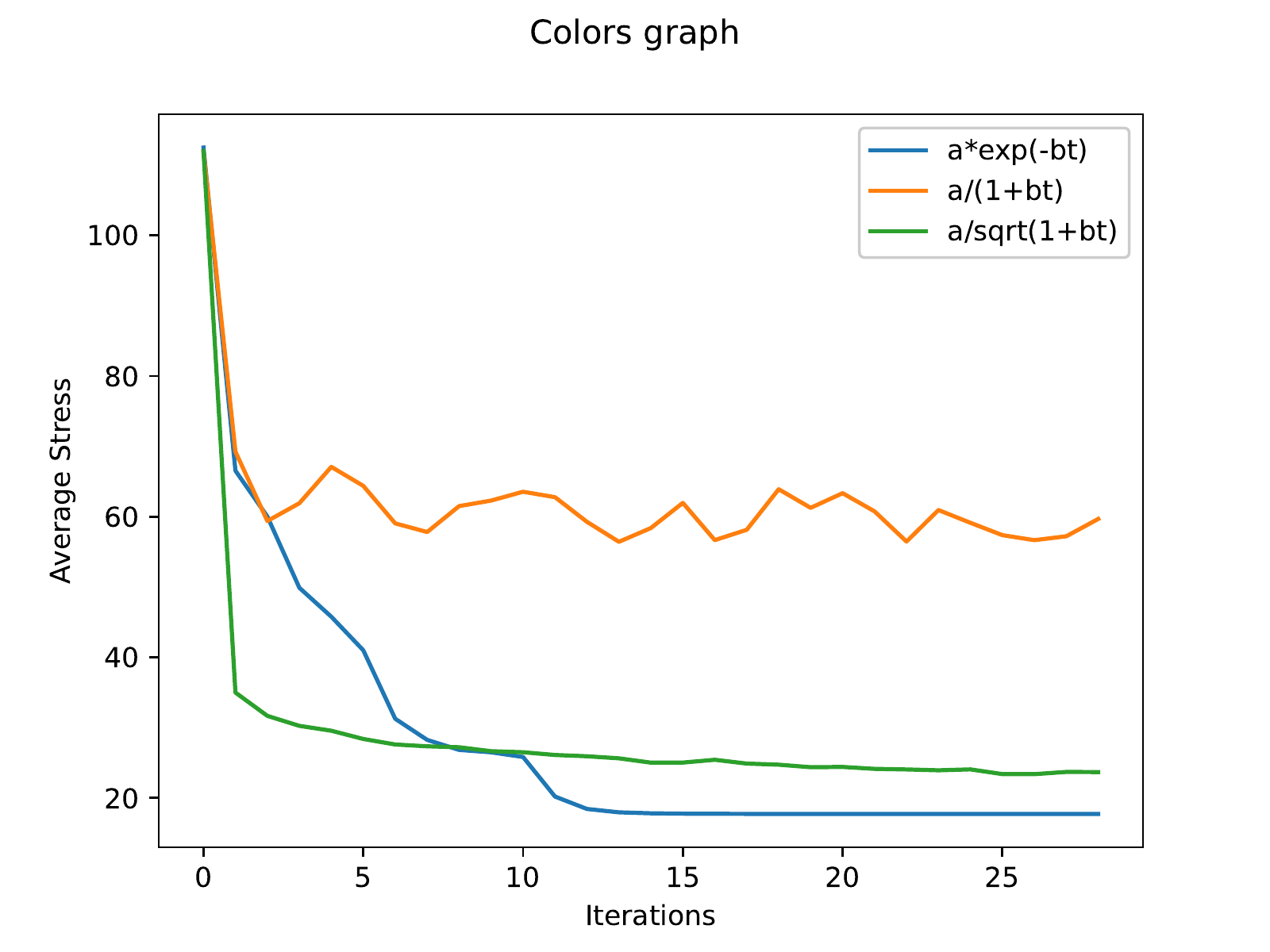}

\caption{Effect of learning rate on various classes of graphs (average over 15 runs each). Graphs used are a 10x10 grid (top left), 50 node random trees (top right), the Les Mis graph~\cite{knuth1993stanford} (bottom left), and the colors graph (bottom right).}
\label{fig:eta-experiment}
\end{figure}

As expected, the $\Theta(1/t)$ schedule struggles to step out of local minima. It is somewhat surprising  that the $\Theta(1/\sqrt{t})$ schedule appears to achieve lower minima for some classes of graphs; see Fig.~\ref{fig:eta-experiment}. This could be due to the function's larger learning rates allowing the system to avoid local minima.

\subsubsection{Stopping Condition}
\label{sec:iterations}
Gradient descent algorithms terminate either if they converge or if they reach a maximum number of iterations.
The convergence is reached when the change in objective function value is less than some tolerance. 
 However, computing the stress value at each iteration is time consuming and we avoid doing this for SGD. Instead, we measure the max change in pairwise distance per iteration. 

For our web focused application, we primarily investigate the use of fixed number of iterations, although one can select to iterate until convergence under `advanced options.' We set $t_{max} = 20$ using the exponential learning rate described above, after experimenting with different input graphs.
We observe that there is little improvement after 20 iterations; see Fig.~\ref{fig:stress_curves}

\begin{figure}
  \centering
  \hspace{-.5cm}\includegraphics[width=.48\linewidth]{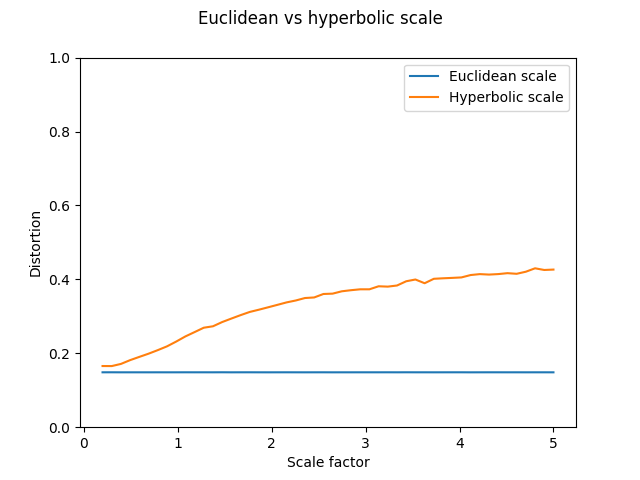}
  \includegraphics[width=.48\linewidth]{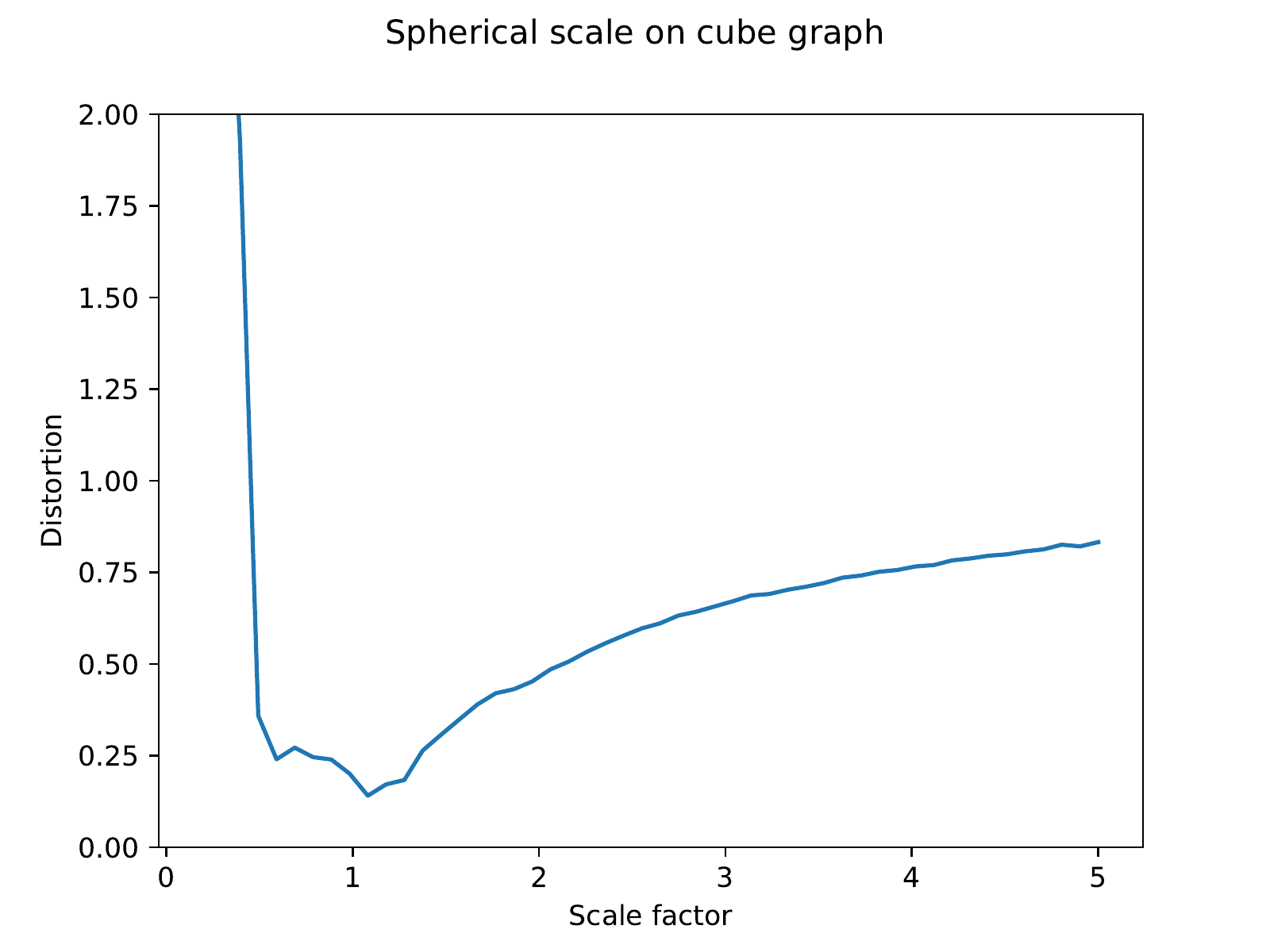}
\caption{Left: Distortion on triangular lattice graph shown in Fig. \ref{fig:unscaledvsscaled}. Hyperbolic space gets worse as the scale increases, but Euclidean can embed the graph with constant error. Right: The effect of scale on the sphere on a cube graph. For this example, there is a noticeable optimum at around $\pi /3$ (note that the diameter of a cube graph is 3).}

\label{fig:scaleinvariance}
\end{figure}

\subsection{Evaluation}
\label{sec:evaluation}

Similar to SGD for Euclidean space, we see similar improvements in time and quality using SGD in hyperbolic space. Experiments were conducted using a desktop machine with an Intel Core i7-3770 CPU @ 3.40GHz x 8 processor, 32 GB of memory, and NVidia GeForce gt 640 graphics running Ubuntu 20.04.3 LTS. Both the GD and SGD algorithms are implemented in Python, making use of the Numpy, Graph-tool, and Numba libraries.

As mentioned in section~\ref{sec:iterations}, while the overall complexity of SGD is no different than GD, the run time is significantly faster; see Fig~\ref{fig:random-effects}. We conduct this experiment by generating a single random graph on $n$ nodes, then computing an embedding using the classic GD and SGD, and recording the average time over 30 runs. Each graph of $n$ nodes has $3n$ edges selected at random. At 500 nodes, GD takes over a minute but SGD takes only about 1.5 seconds. 

\begin{figure}
  \centering
  \includegraphics[width=.48\linewidth]{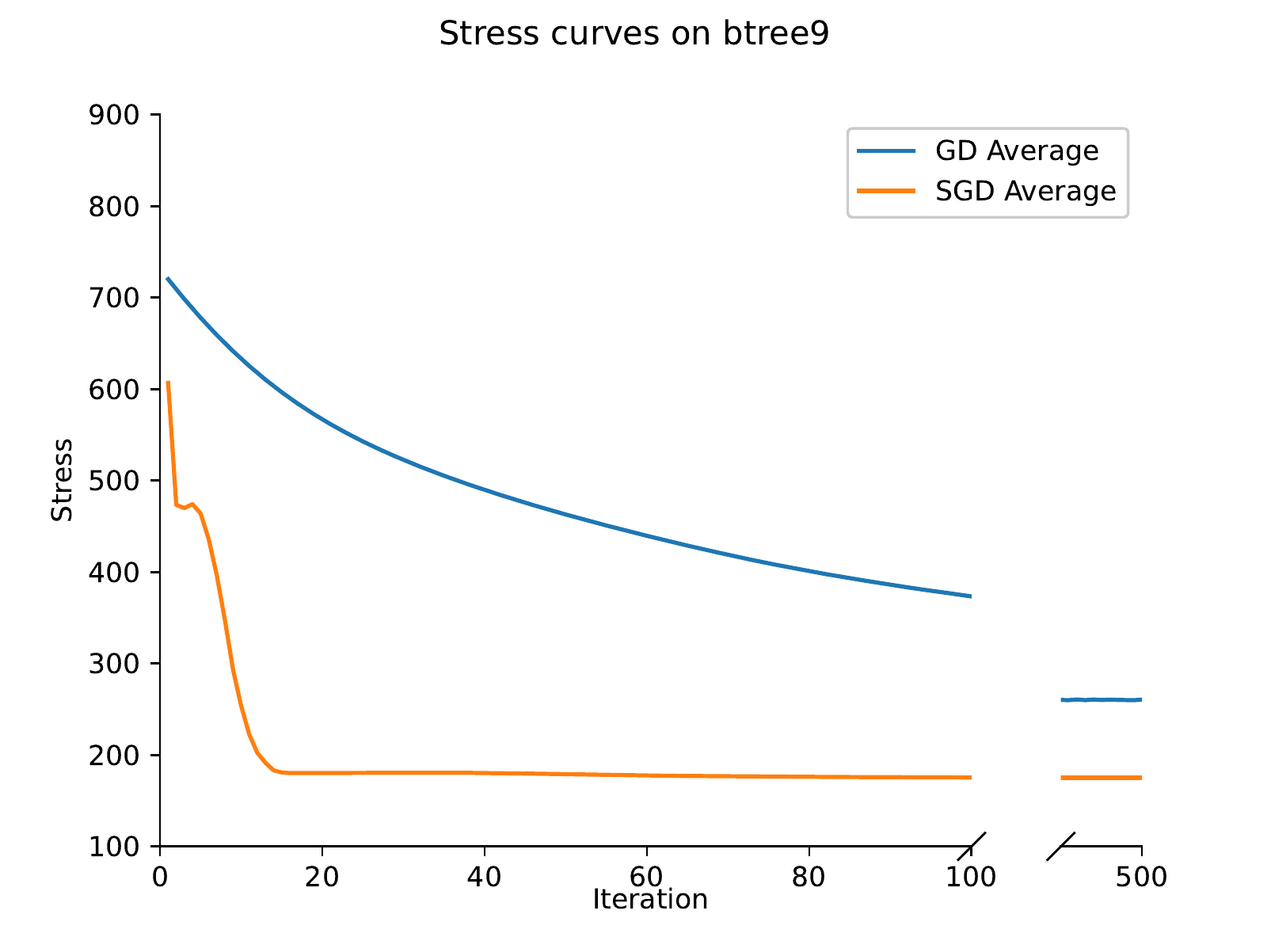}
    \includegraphics[width=.48\linewidth]{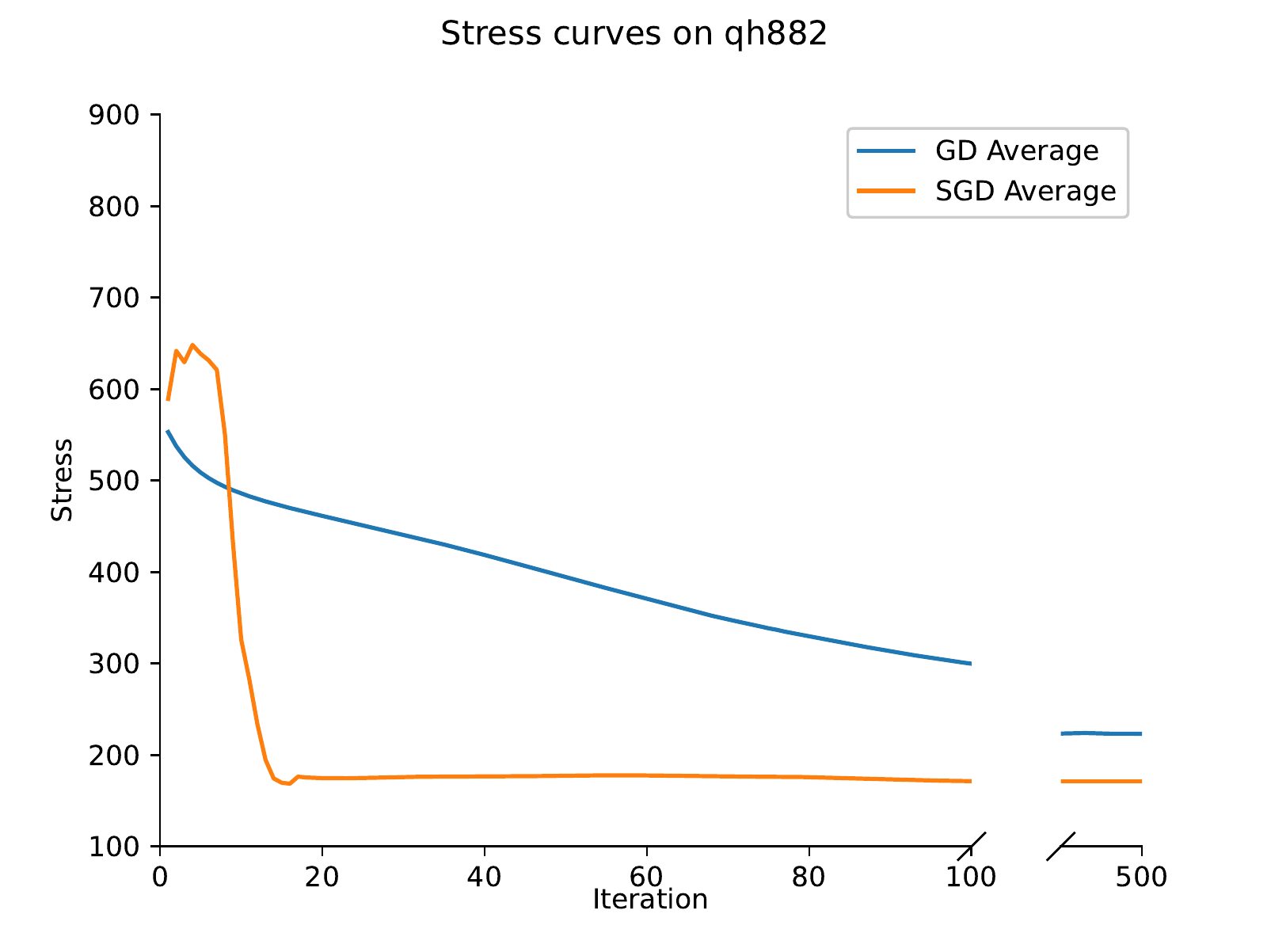}
    
   \includegraphics[width=.48\linewidth]{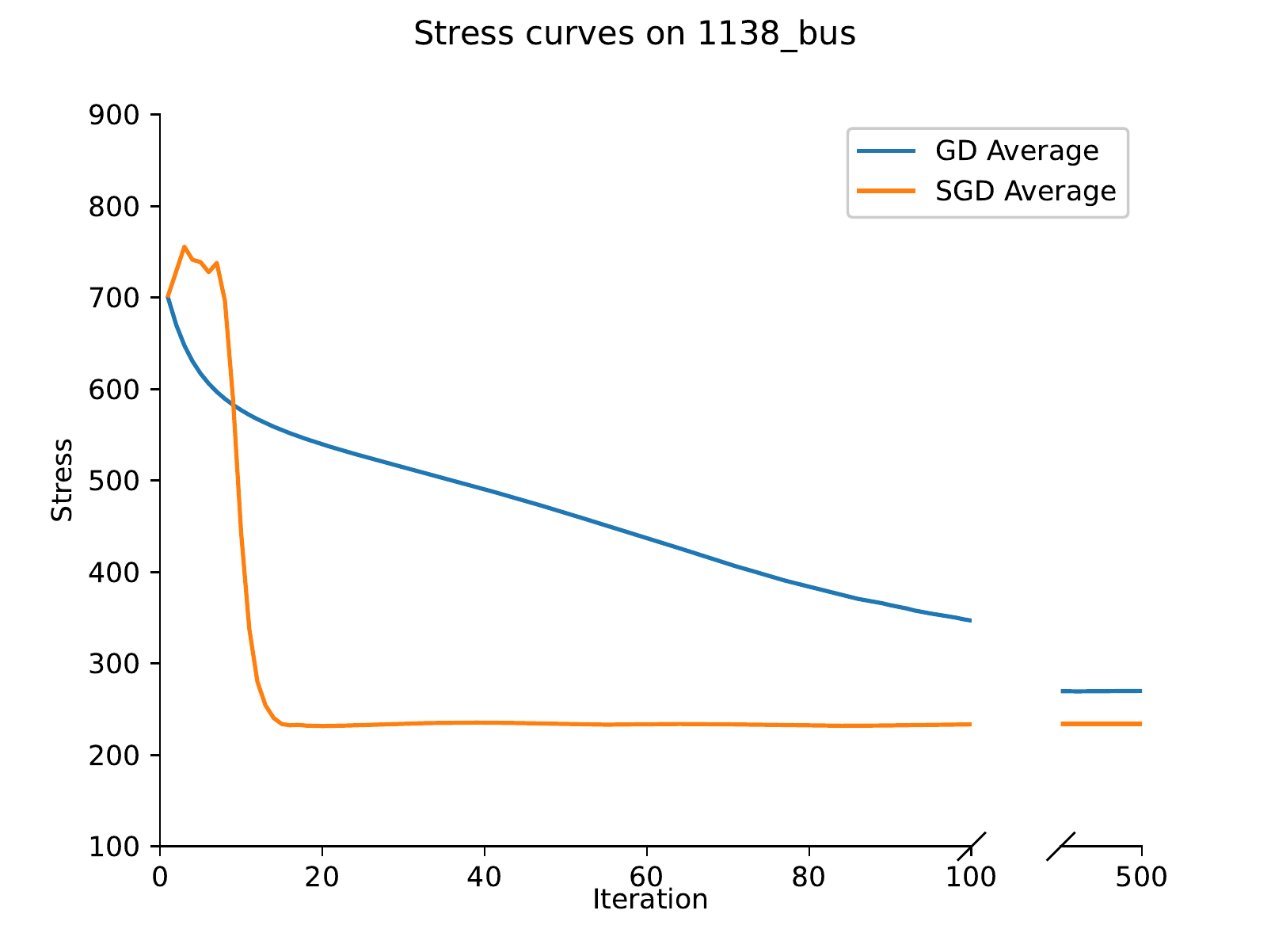}
    \includegraphics[width=.48\linewidth]{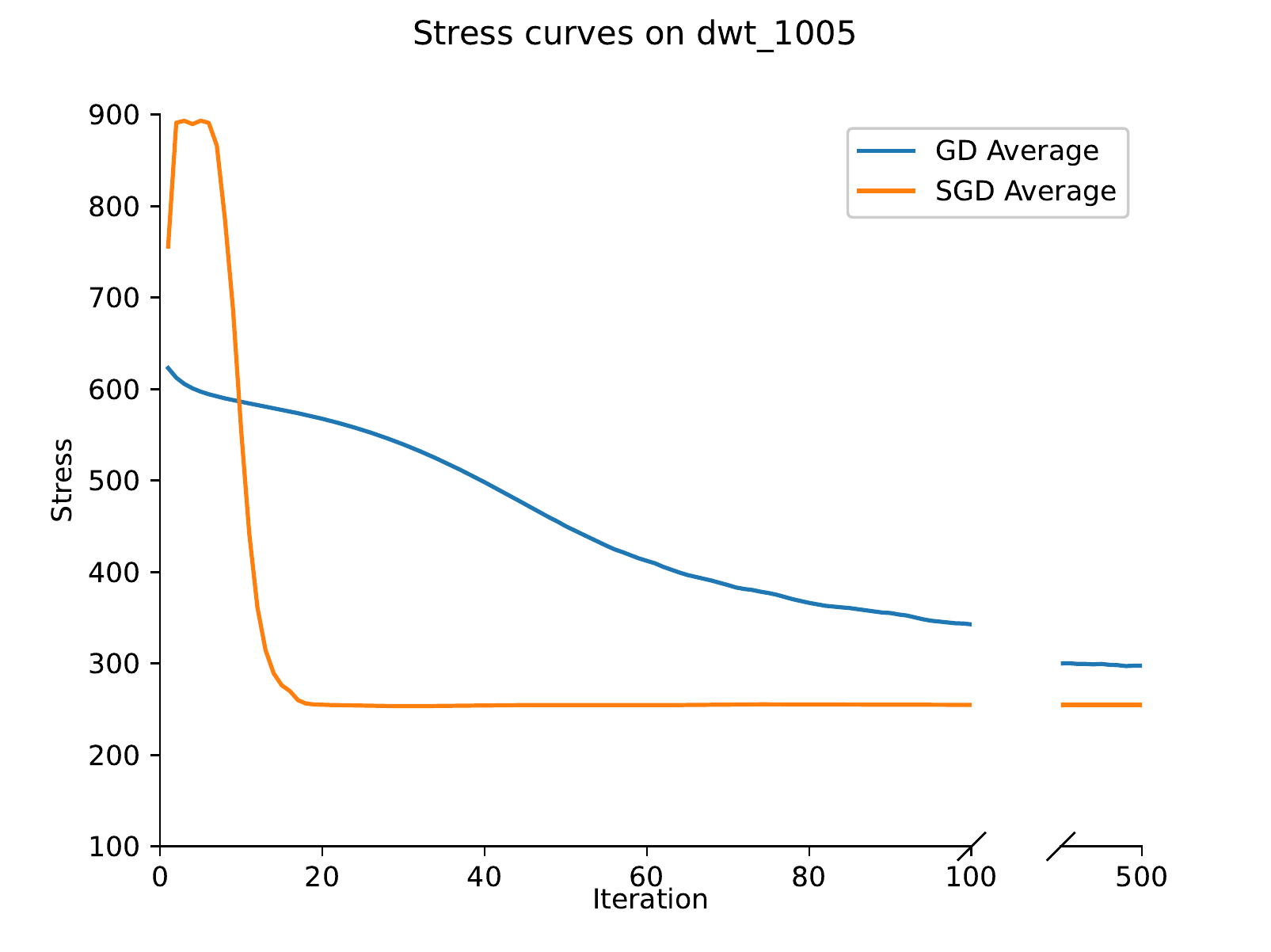}
\caption{Average stress plots of GD and SGD. Initial stress values are omitted.}
\label{fig:stress_curves}
\end{figure}

Consistent with the findings in Euclidean space, hyperbolic SGD also performs better than GD in regards to quality; see Fig.~\ref{fig:stress_curves}. We show a selection of 4 graphs from the sparse matrix collection~\cite{DBLP:journals/toms/DavisH11} and plot the stress minimization curves as each algorithm proceeds. Often just a few iterations of SGD is enough to `untangle' the layout and the curve often bottoms out quite quickly.

\subsubsection{Comparison across geometries}
\begin{figure}

  \includegraphics[width=.48\linewidth]{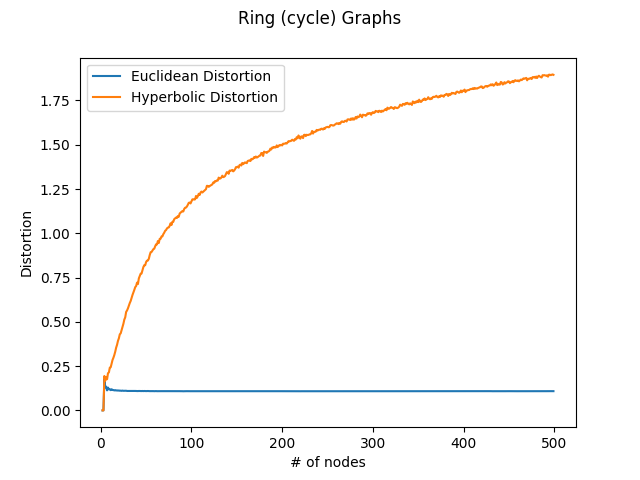}
  \includegraphics[width=.48\linewidth]{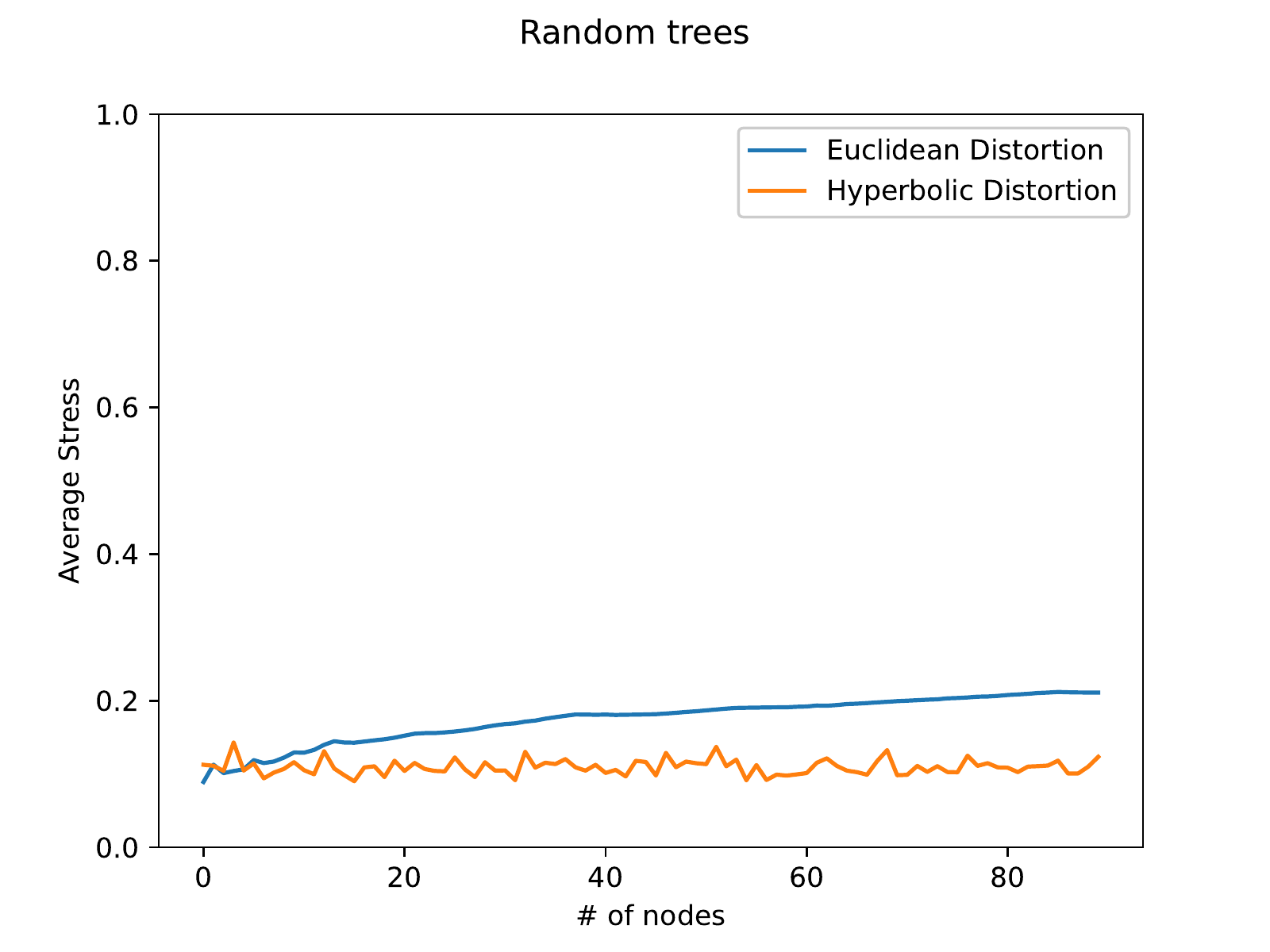}

\caption{Euclidean and hyperbolic embedding distortion on rings (left) and trees (right). It can be seen that the number of nodes in a ring in Euclidean space does not matter, but distortion gets worse with size of the ring in hyperbolic space. The inverse is true for trees, they can be embedded with constant distortion in hyperbolic space but not Euclidean.}
\label{fig:distortions}
\end{figure}

The stress function (Eq.~\ref{eq:stress_def}) offers a natural  evaluation for the quality of a graph embedding, based on how well pairwise distances in the embedding match the corresponding graph distances.
While the quadratic term in the stress equation makes it suitable for gradient descent, it also makes it somewhat disingenuous to compare across different graphs and different embedding spaces. 
With this in mind, we measure how well a graph embedding captures the underlying graph structure using the related distortion measure; see~\cite{sala2018representation}.

\begin{equation}
\label{eq:distortion_def}
\mathit{Distortion} = \frac{1}{{|V|\choose2}}\sum_{i,j}\frac{\left| \Vert X_i -X_j \Vert - d_{ij} \right|}{d_{ij}}
\end{equation}
Similar to stress, perfectly capturing all pairwise distances will result in distortion of $0$. 
%
It has been shown that some classes of graphs, such as trees, can be embedded in the hyperbolic plane with lower distortion value than 
in Euclidean
space~\cite{krioukov2010hyperbolic,blasius2018efficient}. We numerically demonstrate that the SGD algorithm for H-MDS achieves
lower distortion values for trees. 
It has also been shown that some classes of graphs, such as cycles, can be embedded in the Euclidean plane with constant distortion but cannot be embedded with constant distortion in the hyperbolic space. 
We demonstrate 
both of these properties  for trees and cycles; see Fig.~\ref{fig:distortions}.

The ability to compare graph embeddings in various spaces (Euclidean, Spherical and Hyperbolic) creates an interesting application of MDS.
Similar to how Zhou and Sharpee detect the geometry of a dataset~\cite{zhou2021hyperbolic}, we can determine which of the three consistent geometries is best suited for a given graph, by performing Euclidean MDS, Spherical-MDS and H-MDS, and comparing their corresponding distortion values. Table~\ref{tab:geometry-compare} 
shows the distortion values for several classes of graphs for Euclidean space, Spherical space and Hyperbolic space. The cube graph (and other graphs that correspond to 3D platonic solids) embeds best in spherical space. Lattices (as well as paths and cycles) embed best in Euclidean space. Trees (and other hierarchies) embed best in hyperbolic space.

\begin{figure}
    \centering
    \includegraphics[width=0.48\linewidth]{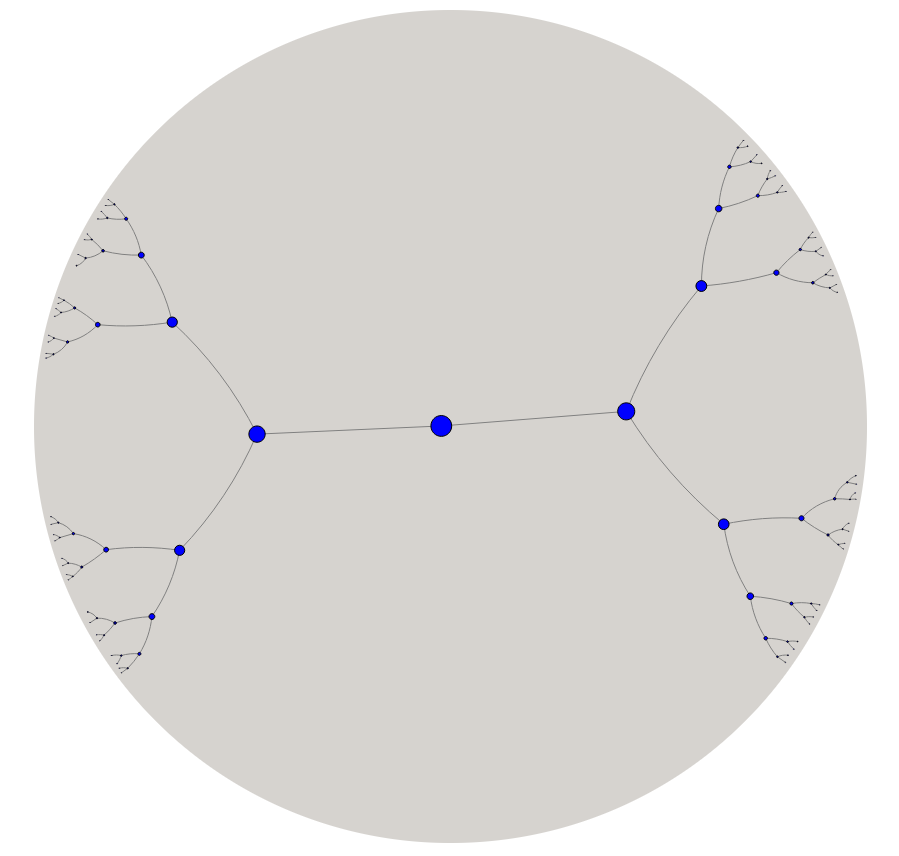}
    \includegraphics[width=0.48\linewidth]{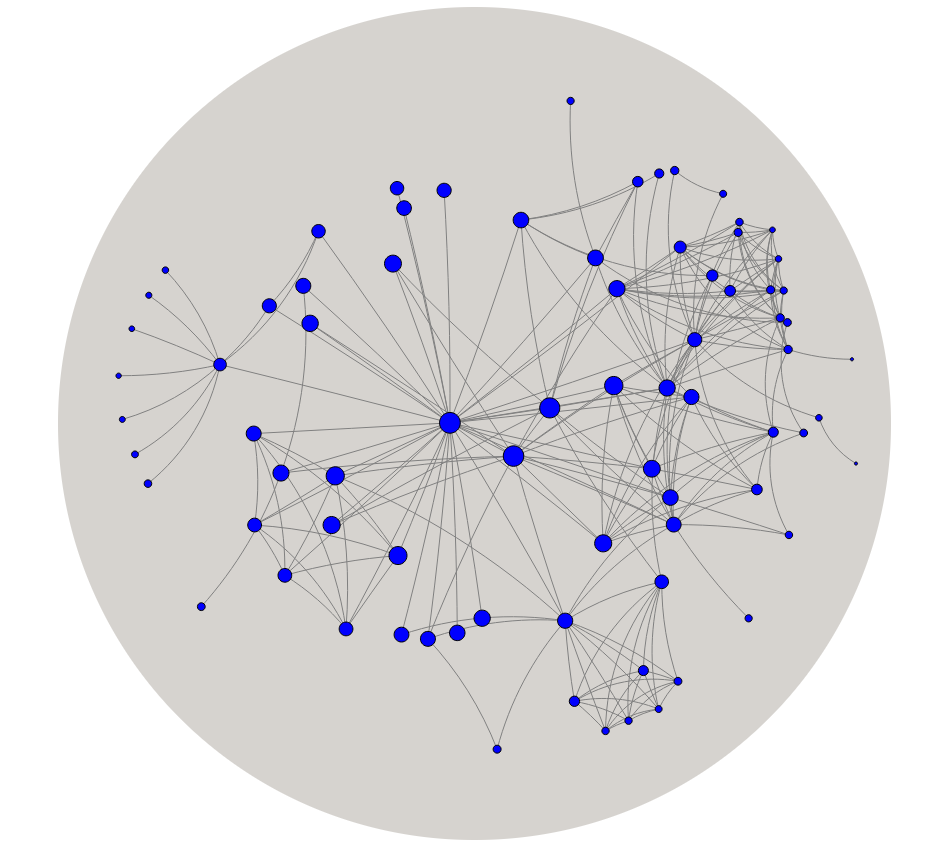}
    \caption{A full binary tree (left) and the Les Mis~\cite{knuth1993stanford} graph (right), an example of a small complex (social) network.}
    \label{fig:hmds-example}
\end{figure}



\subsubsection{Scale Invariance}
\label{sec:scale}

It is known that the Euclidean MDS is invariant to scale. That is, given a distance matrix and its corresponding embedding by MDS, if one scales all distances by the same scalar and applies MDS to the scaled distances, the achieved embedding should be the scaled version of the initial one. However, this property does not hold for spherical-MDS (S-MDS) and H-MDS. 
%
This can perhaps be most intuitively seen by looking at the non-Euclidean analogues of the Pythagorean theorem (assuming unit curvature). \newline
\textit{Euclidean}: $a^2 + b^2 = c^2,$
\newline
\textit{Spherical}: $cos(a) + cos(b) = cos(c),$
\newline
\textit{Hyperbolic}: $cosh(a) + cosh(b) = cosh(c).$
\newline
While we can multiply both $a$ and $b$ by the same constant $k$ to obtain $k^2 c^2$ in Euclidean space, the same property does not hold for hyperbolic and spherical spaces.

So then, our objective function for H-MDS becomes 
$$\textit{Stress} = \sum_{i<j}w_{ij}(\textit{gdist}(X_i, X_j) - \alpha d_{ij})^2,$$
where the $\textit{gdist}((X_i, X_j))$ is the geodesic distance in hyperbolic space between nodes $X_i$ and $X_j$.

Spherical space is even more problematic when considering embedding scales, as for any given radius of the sphere, the maximum distance that one can achieve on the sphere is finite (rather than infinite in Euclidean and hyperbolic space). This leads to
a natural heuristic scale value: $\alpha = \frac{\pi}{d_{max}}$, where $d_{max}$ is the diameter (longest shortest path) of the graph. This normalizes $d$ to a maximum distance of $\pi$, which is the longest distance possible on the unit sphere. 

\begin{figure}
  \includegraphics[width=.48\linewidth]{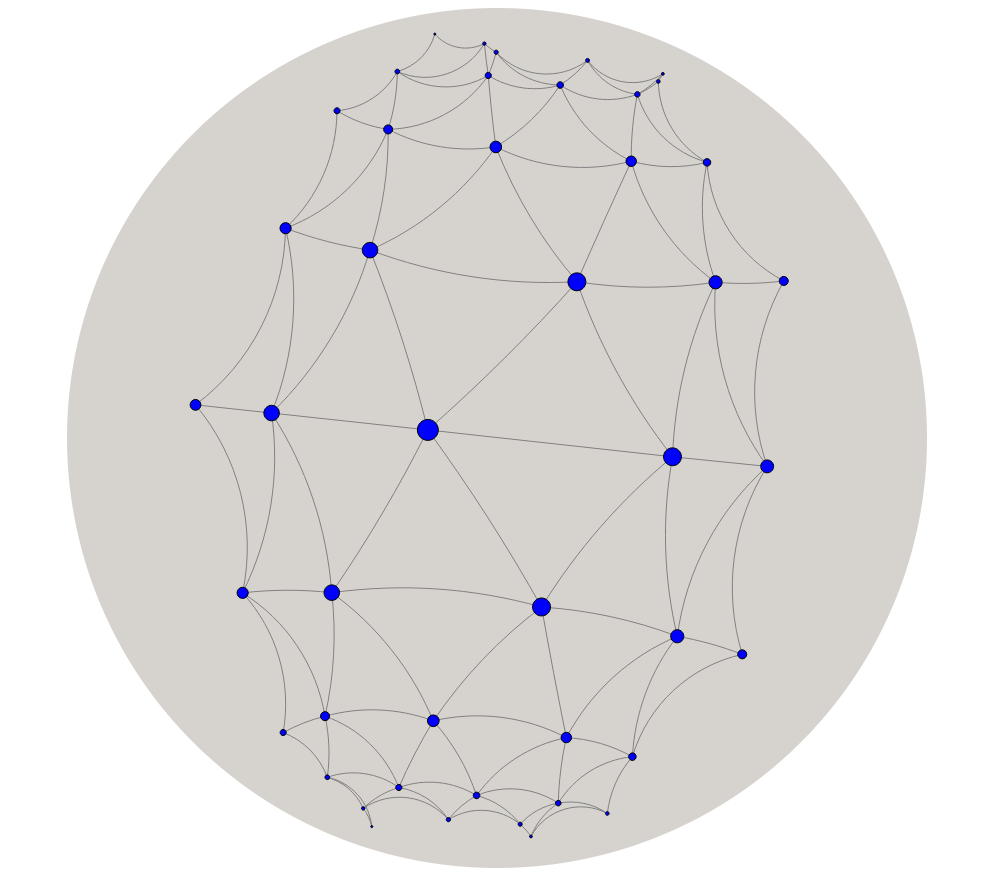}
  \includegraphics[width=.48\linewidth]{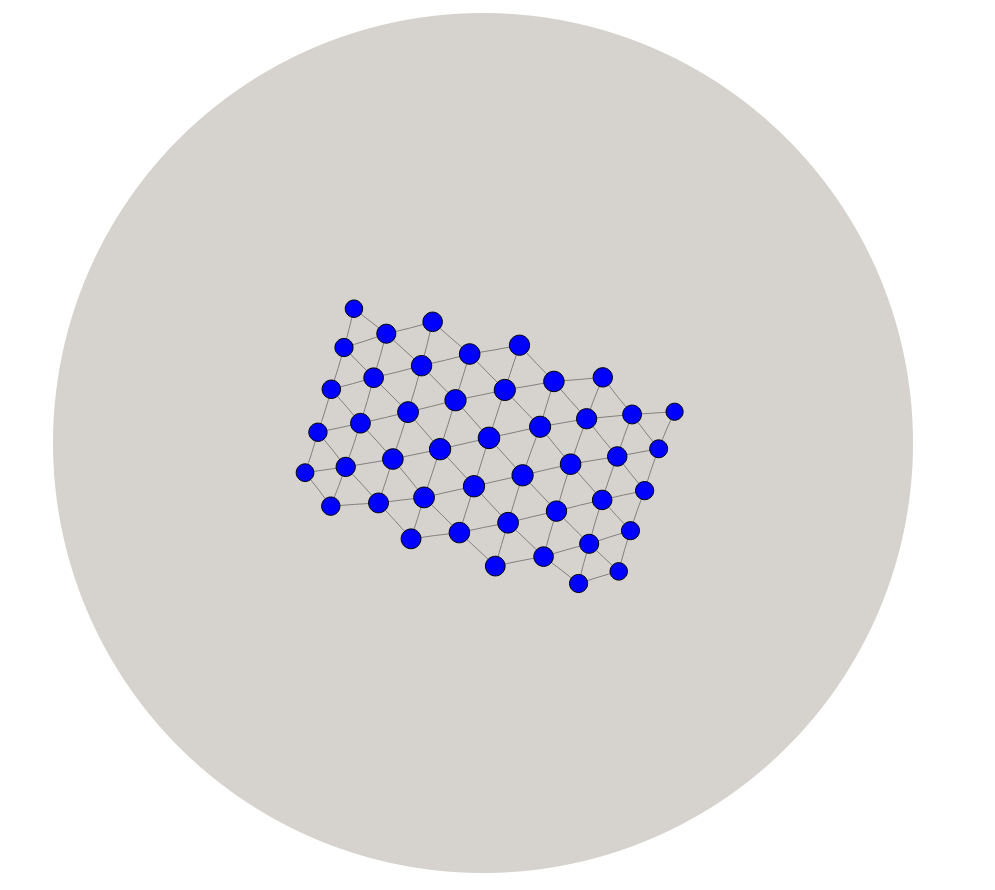}

\caption{Triangular lattice with scaling factor $\alpha = 1$ (left) and optimized $\alpha = 0.22$ (right).}
\label{fig:unscaledvsscaled}
\end{figure}

In the hyperbolic space, although one can achieve arbitrarily large distances, similar to the S-MDS, scaling the data or considering a different hyperbolic radius can drastically affect the embedding. Thus, there is a need to find an appropriate scaling parameter $\alpha$ for which the achieved embedding best captures the underlying graph distances.
If $\alpha$ is very small, the layout occupies a small fraction of the hyperbolic space, resulting in an embedding that is similar to Euclidean space, and thus does not capture the focus+context effect.
If $\alpha$ is large, then most of the graph is located at the periphery, making it hard to see.
We can find a good scaling parameter for any given graph using binary search for the value of $\alpha$ that achieves lowest embedding distortion and this is indeed an available option under `advanced options.' 
We show an example of an optimized $\alpha$ compared to a naive $\alpha = 1$; see Fig.~\ref{fig:unscaledvsscaled}.
By default we set $\alpha = \frac{10}{d_{max}}$, where $d_{max}$ is the length of the longest shortest path in the graphs. This caps the largest distance to a hyperbolic unit length of 10 and the resulting embeddings tend to capture the focus+context effect and do not place large parts of the graph near the periphery.

\begin{table}[h]
 \caption{Distortion on small graphs across geometries. Averaged over 10 runs.}\vspace{1ex} 
 \label{tab:geometry-compare}
 \scriptsize
 \centering 
   \begin{tabular}{|r|r|r|r|}
   \hline
     Graph & Spherical  & Euclidean & Hyperbolic\\
   \hline
   \hline
     Cube & 0.1296  & 0.2437 & 0.2645\\
   \hline
     Lattice & 0.2421 & 0.1486 & 0.2306\\
   \hline
     Tree & 0.1944 & 0.1284 & 0.0682\\
    \hline
     
   \end{tabular}
\end{table}

\section{Discussion, Limitations and Future Work}

\begin{figure}

 \centering
  \includegraphics[width=.9\linewidth]{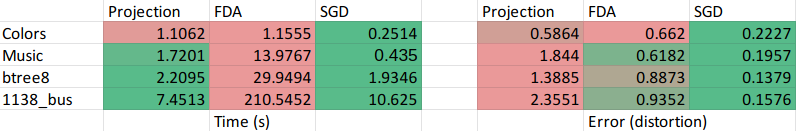}

\caption{Average time in seconds (left) and average distortion value (right) on the listed graphs for each of three methods presented in the paper.}
\label{fig:exp3}
\end{figure}

We described three methods for visualizing graphs in hyperbolic space, which are illustrated in Fig.~\ref{fig:teaser}. 
We present a small-scale comparison of the three approaches by comparing time and distortion values; see Fig.~\ref{fig:exp3}.
%
%
The projection-based method allows us to show any 2D Euclidean graph representation in hyperbolic space, where we can take advantage of the `focus+context' properties of the space while still relying on standard map interactions. Related work has been limited to standard node-link representations, but this method can be applied to {\it any} graph visualization metaphor, which we show with GMaps, MapSets, BubbleSets, and LineSets.
The method currently relies on Lambert azimuthal projections and the Poincar\'e disk model. We have not yet explored other projections or the Beltrami-Klein model.
Finally, this method does not fully take advantage of the underlying geometry of the space.

The inherent distortion of shapes and angles introduced when using the inverse Lambert projection to the
hyperbolic plane implies that at some threshold the outer regions of the layout become
too distorted to be of use. This is already apparent in the MusicLand example
from GMap as shown in Fig.~\ref{fig_music}, with around 250 nodes. 
Even though our method can handle larger graphs, it is clear that larger graphs pose additional challenges. A multi-level representation of the graph might be useful to provide `semantic zooming' where we start with a high level overview of the graph and zooming in brings up more details, following Schneiderman's mantra (overview first, zoom and filter, details on demand). 

When moving through a curved space, an inherent property causes an observer to incur rotation. This could be desirable, as it gives several different
perspectives on the same layout, but it could potentially be confusing when navigating large maps.
Specifically, moving the layout in the Poincar\'e disk, incurs a rotation in the layout (clockwise or counter-clockwise): consider translating a layout some fixed distance up, the same distance to the right, then again down, and back to the left. In 2D Euclidean geometry, the layout would be identical after these transformations, while in the Poincar\'e disk (and hyperbolic geometry in general) this causes a 90-degree rotation.
An orientation correcting transformation could be applied after translating the layout, but in our prototype we only provide the `reset button,' which restores the original layout.

The force-directed method  utilizes the geometry of hyperbolic space, but is not as efficient as our projection-based method. The underlying Kamada-Kawai algorithm is
already rather computationally expensive, due to the all-pairs shortest
path calculations  
and many tangent plane computations.
There are several scalable force-directed algorithms for Euclidean space, which can be adapted to the hyperbolic setting, but this remains as future work.

Our third method 
lays out a graph directly into the hyperbolic plane using H-MDS, a generalization of multidimensional scaling. The algorithm is implemented, fully functional and available online on GitHub.  
In order to optimize the stress function of H-MDS we employed stochastic gradient descent, which not only significantly improves the runtime, but often finds a better minimum when compared to gradient descent.
H-MDS also requires an all-pairs-shortest-paths computation, but  relatively few iterations. Adapting a sparse approximation method for graphs in which the pre-processing is prohibitively expensive could be a direction for future work.

In this work we visualized the hyperbolic space by using the Poincar\'e disk model,
as it provides the look and feel of hyperbolic space. Other models such as the Beltrami-Klein model or Poincar\'e half-plane model may provide additional benefits for visualization. 

While we have addressed the computational scalability of hyperbolic layouts with hyperbolic SGD, visual scalability remains an open problem. Recent work has pointed to limitations on hyperbolic graph embeddings~\cite{DBLP:conf/chi/DuCLXT17,eppstein2021}, but it is also known that some graphs can be embedded in hyperbolic space with lower error~\cite{krioukov2010hyperbolic} Determining whether a lower distortion in a geometry corresponds to better task support remains a promising direction for future work.

As discussed in Section~\ref{sec:scale}, scaling is crucial for hyperbolic and spherical embeddings and a robust algorithm to efficiently determine the correct scaling parameter for H-MDS and S-MDS is needed. We provide an efficient heuristic for setting the scale and provide a more computationally expensive method to find a scale  that minimizes distortion, but a closed form solution remains an open problem.


A possible promising application for hyperbolic/spherical visualization are
virtual reality and augmented reality,
as prior work seems to have only considered spherical space in this context~\cite{kwon2016study}.
%

A potentially interesting question is how the hyperbolic geometry may change the layout's aesthetic properties. Wang {\it et al.} optimize edge orientation to better facilitate navigation tasks on graphs using a fish-eye lens~\cite{DBLP:journals/tvcg/WangWZSFSCD19}. Perhaps optimizing over additional aesthetic criteria could improve  the readability of hyperbolic graph layouts.

%


\bibliographystyle{abbrv-doi}

\bibliography{hyperbolic_graphs}
\end{document}